\documentclass[sigconf,10pt]{acmart}

\usepackage[english]{babel}
\usepackage{blindtext}
\usepackage{subcaption}
\usepackage{algorithmicx}
\usepackage[ruled]{algorithm}
\usepackage[noend]{algpseudocode}
\usepackage{paralist}
\renewcommand\footnotetextcopyrightpermission[1]{} % removes footnote with conference info
\setcopyright{none}

\usepackage{enumitem}
\setlist{nolistsep}

\usepackage{cleveref}
\crefname{section}{}{\S\S}
\crefdefaultlabelformat{\S#2#1#3}

\acmDOI{}

\acmISBN{}

\acmPrice{}

\newcommand{\sys}{Flash\xspace}
\newcommand{\parabf}[1]{\medskip\noindent\textbf{#1}}
\newcommand{\parait}[1]{\medskip\noindent\textit{#1}}

\begin{document}
\title{\sys: Efficient Dynamic Routing \\for Offchain Networks}

\author{Peng Wang}
\affiliation{
  \institution{City University of Hong Kong}
}
\email{pewang4-c@my.cityu.edu.hk}

\author{Hong Xu}
\affiliation{
  \institution{City University of Hong Kong}
}
\email{henry.xu@cityu.edu.hk}

\author{Xin Jin}
\affiliation{
  \institution{Johns Hopkins University}
}
\email{xinjin@cs.jhu.edu}

\author{Tao Wang}
\affiliation{
  \institution{City University of Hong Kong}
}
\email{twang55@um.cityu.edu.hk}

\begin{abstract}
Offchain networks emerge as a promising solution to address the scalability
challenge of blockchain. Participants directly make payments through a network
of payment channels without the overhead of committing onchain transactions.
Routing is critical to the performance of offchain networks.
Existing solutions use either static routing with poor performance or dynamic
routing with high overhead to obtain the dynamic channel balance information.
In this paper, we propose \sys, a new dynamic routing solution that leverages
the unique
characteristics of transactions in offchain networks to strike a better tradeoff
between path optimality and probing overhead.
By studying the traces of real offchain networks, we find that the payment sizes
are heavy-tailed, and most payments are highly recurrent.
\sys thus differentiates the treatment of elephant payments from mice payments.
It uses a modified max-flow algorithm for elephant payments to find
paths with sufficient capacity, and strategically routes the payment across
paths to minimize the transaction fees.
Mice payments are directly sent by looking up a routing table with a few
precomputed paths to reduce probing overhead.
Testbed experiments and data-driven simulations
show that \sys improves the success volume of payments by up to 2.3x compared to
the state-of-the-art routing algorithm.
\end{abstract}

\maketitle

\section{Introduction}
\label{sec:introduction}

Blockchain is the fundamental infrastructure to support a new generation of
decentralized Internet applications. It has enabled many innovations from
decentralized cryptocurrencies to smart contracts~\cite{nakamoto2008bitcoin, wood2017ethereum}.
\emph{Scalability} is the primary challenge for blockchain to support
decentralized applications at the Internet scale~\cite{WW19,nakamoto2008bitcoin,
gilad2017algorand, luu2016secure, poon2016bitcoin}. As a concrete example,
Bitcoin only processed fewer than 20 transactions per second at peak from
November 2018 to January 2019~\cite{tps}, whereas Visa was reported to process
more than 47,000 transactions per second at peak during the 2013 holiday seasons
\cite{poon2016bitcoin}.

Blockchain is intrinsically difficult to scale because it aims to achieve a
global consensus between all participants, which involves complex protocols to
consistently replicate all state changes to every participant. Despite various
efforts to improve the efficiency and reduce the overhead of blockchain
protocols~\cite{WW19, zamani2018rapidchain}, their performance is ultimately limited by the
network bandwidth and processing capability of the participants to replicate
state changes.

\emph{Offchain networks} (or payment channel networks) emerge as one of the most
promising solutions to solve this dilemma~\cite{gilad2017algorand,
luu2016secure, poon2016bitcoin}. Two participants can use a bidirectional
payment channel to make \emph{multiple} payments with each other, without the need to
commit every transaction to the blockchain. The blockchain is only involved when
the participants set up and tear down the payment channel, or when they have
disagreements on the transaction results. A network of payment channels form an
\emph{offchain network}, and allows two participants without direct channels to
send payments via a multi-hop path. A transaction can be safely committed in an
offchain network, as soon as the participants on the payment path achieve an
agreement with a multisignature contract such as a Hashed Timelock Contract
(HTLC)~\cite{poon2016bitcoin}. This obviates the need to consistently commit the transaction
to \emph{every} participant on the blockchain. As a result, offchain networks
have the potential to significantly improve the transaction throughput and
reduce the transaction latency of blockchain. Examples including Lightning
Network~\cite{lightningnetwork}, Raiden Network~\cite{raiden}, and
Ripple~\cite{ripple} are increasingly being adopted and used in practice.

Routing is critical for offchain networks to fulfill their promise. Efficient
routing can successfully deliver most payments in an offchain network,
minimizing the operations on the blockchain. While routing is a well-studied
problem in computer networking, there is a key difference between an offchain
network and a computer network. In a computer network, each link has static
capacity and the capacity does not change as the link sends packets. However, in
an offchain network, the initial balance of a payment channel (i.e. channel
capacity) is deposited by the participants during the channel setup, and the
balance is updated after \emph{every} payment in the channel. As a result,
offchain networks are more dynamic than computer networks.

At a high level, there are two major classes of protocols for network routing.
The first uses static routing where the path for each flow is fixed after
(periodical) path discovery. Many traditional routing protocols such as OSPF
and IS-IS fall into this class. Static routing guarantees reachability, and is
typically used when the network topology and traffic are mostly static, or if
the network capacity is abundant and the performance is not a concern. Early
routing protocols for offchain networks, such as
Flare~\cite{prihodko2016flare}, SlientWhispers~\cite{moreno2017silentwhispers}
and SpeedyMurmurs~\cite{roos2018settling}, leverage this approach, but they
suffer from low transaction throughput, because they do not consider dynamic
channel balance in offchain networks. The limitation of static routing
motivates the second class of protocols that use dynamic routing, where the
path for each flow or flowlet is dynamically updated based on real-time
network characteristics. Many emerging solutions in datacenters and
inter-datacenter WANs~\cite{AEDV14,VPAT17,WXNH16,JKMO13,HKMZ13} fall into this class.
Spider~\cite{sivaraman2018routing} applies dynamic routing to offchain
networks and achieves higher performance than earlier static protocols for
blockchain.

Dynamic routing, however, is not a panacea. It is well-known that there exists
a trade-off between path optimality and probing overhead. Using an optimal
path comes at the cost of probing the network in the first place. This is
especially true for offchain networks, as the channel balance changes after
each payment, and one needs to pay the probing overhead for every single
payment if an optimal path was to be chosen.

Classical solutions in computer networking suggest to
strike a balance between path optimality and probing overhead by differentiating
the treatment of elephant flows from that of mice flows~\cite{ARRH10,CMTY11}. A small
number of elephant flows are dynamically scheduled on different paths for high
performance, and the vast mice flows are randomly mapped to static paths for low
probing overhead. Realizing the idea of elephant flow routing in offchain
networks is challenging for at least two reasons.

\begin{itemize}
		\item First, we need to understand whether elephant flow routing is suitable for offchain networks. If all payments in offchain networks have
		similar size, then there does not exist elephant payments in the first place.
		Even if so, if mice payments are highly ad hoc, i.e. a sender often
		chooses a different receiver to send a payment, significant probing for mice
		payments can still not be avoided.

		\item Second, we need to design an efficient protocol to satisfy the unique
		requirements of offchain networks. Because offchain networks are more
		dynamic than computer networks, elephant payments need to probe more
		paths aggressively in order to find efficient paths. The problem is
		exacerbated by the distributed nature of offchain networks. Unlike datacenters
		networks and inter-datacenter WANs that have centralized control planes,
		blockchain systems are decentralized, and there is no coordination between
		participants.
\end{itemize}

To address these challenges, we first conduct a measurement study on the payment
traces of two real-world blockchain networks, Ripple~\cite{ripple} and
Bitcoin~\cite{bitcoin}. By analyzing the traces, we find two important
characteristics of cryptocurrency transactions. First, the payment sizes of
transactions exhibit heavy-tailed distributions. Most transactions are small,
while a small number ($<$10\%) of transactions contribute to the most (over
94\%) of the total transaction volume. This demonstrates the existence of
elephant transactions in offchain networks. Second, payments are highly
recurrent. Within a period of 24 hours, over 80\% of transactions happen between
existing pairs of participants. This suggests that most mice payments can
leverage existing paths with no extra probing overhead, instead of discovering
new paths.

Based on these findings, we design \sys, a new dynamic routing solution for
offchain networks that separate elephant payments from mice payments to
achieve high throughput with low probing overhead. \sys uses a modified
max-flow algorithm based on Edmonds-Karp \cite{CLRS09} to probe and find
candidate paths with sufficient capacity for elephant payments, and carefully
distribute the payments over these paths to minimize transaction fees
(collected by intermediate nodes). Mice payments are sent through a few
existing paths if they have already been computed to minimize the need of
probing. Payments are split into trunks to be sent through multiple paths when
needed, since the atomicity of the transactions can be ensured with recent
proposals such as Atomic Multi-Path Payments (AMP)~\cite{amp}.

In summary, we make the following contributions.
\begin{itemize}
	\item We perform a measurement study on the payment traces of real blockchain
	networks to understand the traffic characteristics of cryptocurrency
	transactions.

	\item We design \sys, a new routing protocol for offchain networks that
	separate elephant payments from mice payments to achieve high throughput with
	low probing overhead.

	\item We implement a prototype of \sys on a cluster of commodity servers.
	Testbed experiments and trace-driven simulations show that \sys improves
	the network throughput by up to 2.3x compared to the state-of-the-art
	routing algorithm. The code of \sys and the
	offchain network traces are open source on Github.\footnote{https://github.com/NetX-lab/Offchain-routing-traces-and-code}
\end{itemize}

\section{Background and Motivation}
\label{sec:motivation}

In this section, we first give a brief introduction of offchain networks,
and then motivate our design with our findings in real traces from
Ripple~\cite{ripple} and  Bitcoin~\cite{nakamoto2008bitcoin}.

\subsection{A Primer on Offchain Networks}

\parabf{Payment channels.} Payment channels are a basic building block of
offchain networks. A payment channel is established between
two parties, and allows them to make multiple payments without the need to
commit every payment to the chain. To ensure the offchain security, both
parties maintain a multi-signature address which guarantees that any balance
updates on the channel require mutual agreement. The chain is only involved
when there is a dispute regarding current balance or setting up and tearing
down the channel. By moving payments away from the chain, it reduces
computation and replication overhead, improves transaction throughput, and
lowers latency.
Furthermore, because sending payments over payment channels does not need to
reward ``miners'', payment channels provide competitively low transaction fees
and better support payments.

\begin{figure}[t]
\centerline{
  \includegraphics[width=0.48\textwidth]{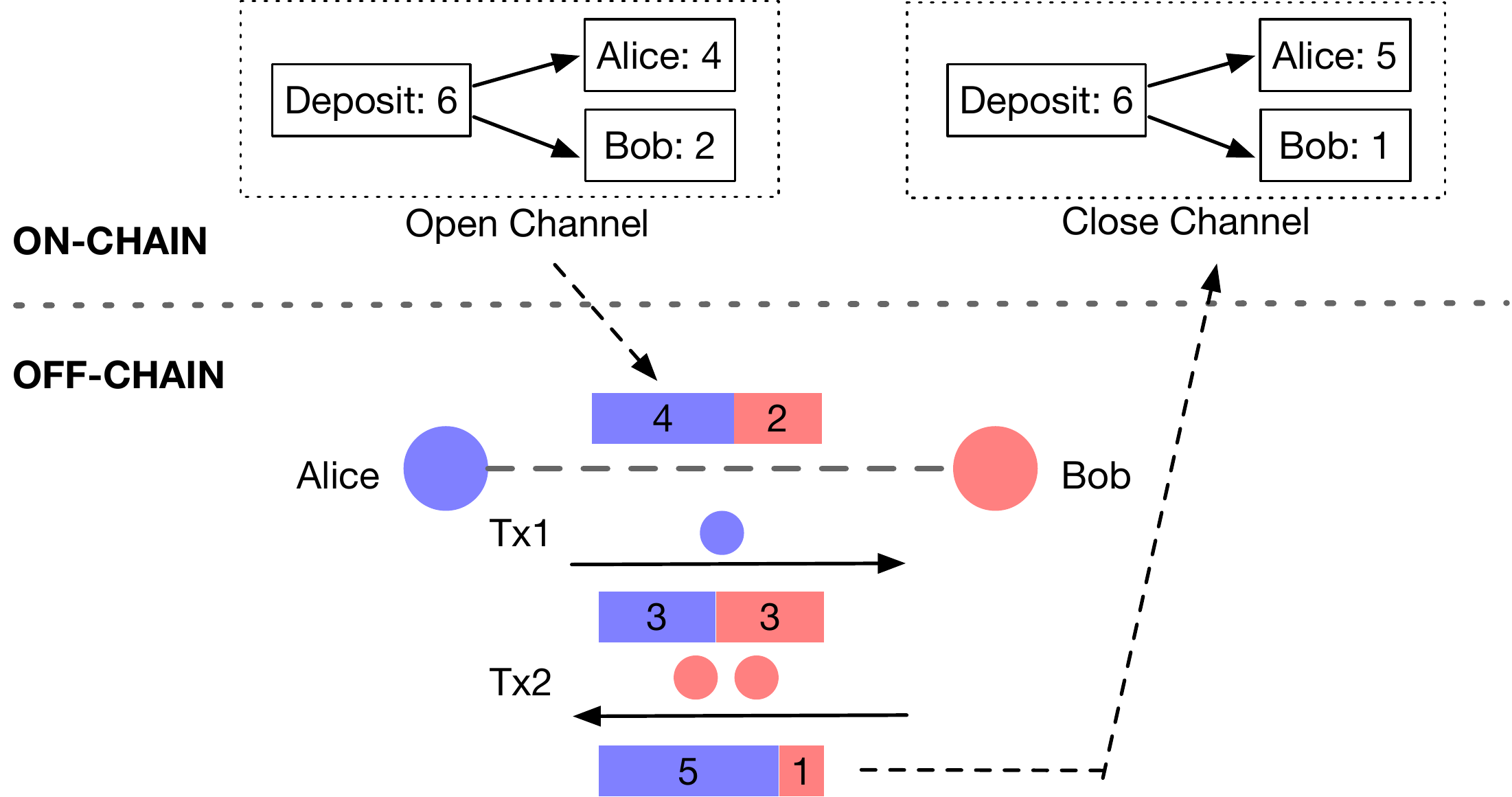}}
\caption{A payment channel between Alice and Bob. Alice and Bod deposit 4 and 2
satoshis respectively to open a channel. Two payments are made off the chain.
Alice
first pays Bob 1 satoshi, and then receives 2 satoshis from Bob. Finally, the
channel is closed by committing the latest state to the chain.}
\label{fig:payment-channel}
\end{figure}

A toy example in Figure~\ref{fig:payment-channel} illustrates the basic
operations of a payment channel with bitcoin as the cryptocurrency. To open a
channel, Alice and Bod jointly make a transaction on the chain, in which they
deposit funds to the channel. The channel is established after the transaction
is committed to the chain. In this example, Alice funds the channel with 4
satoshis and Bob with 2 satoshis (Satoshi is the smallest unit of bitcoin). At
this point, the balance---which limits the maximum amount of bitcoin one party
can send to the other---becomes 4 satoshis for Alice and 2 satoshis for Bob.
The balance of each party is then updated after each successful transaction
executed based on mutual agreement. Thus if Alice pays Bob 1 satoshi, both
would have a balance of 3 satoshis. As long as the channel remains open and
the payment from one to the other does not exceed its balance, Alice and Bob
can repeatedly make any number of transactions. Finally, Alice and Bob can
choose to close the channel if no more transactions are needed. The final
state of the channel is committed to the chain, and both can withdraw their
balances.

\begin{figure}[t]
\centerline{\includegraphics[width=0.48\textwidth]{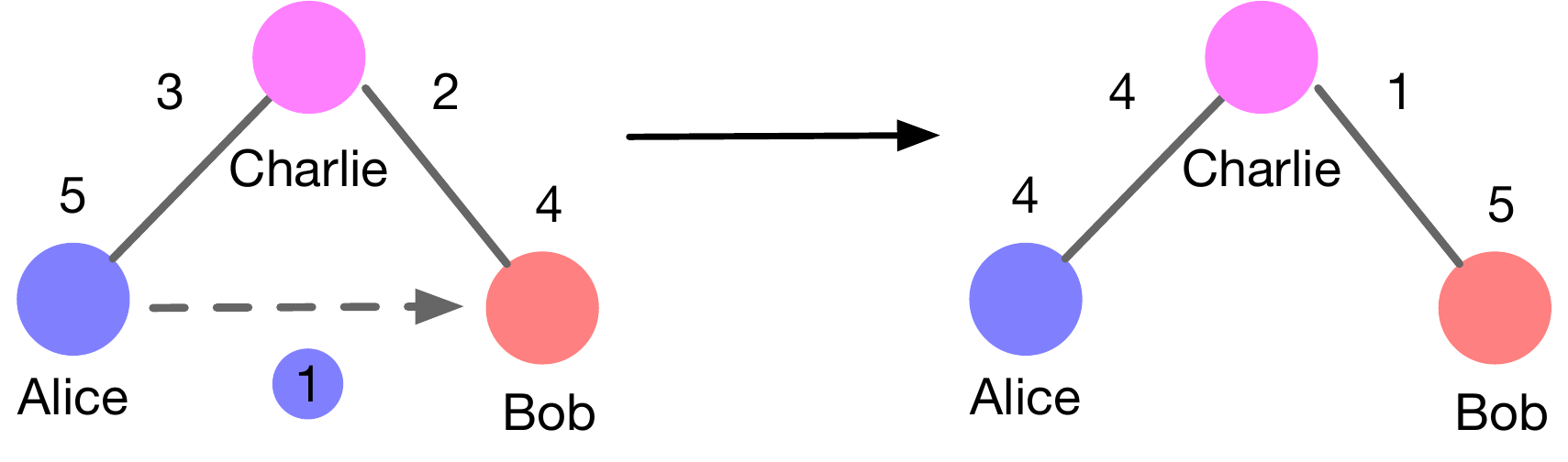}}
\caption{An indirect payment on a payment channel network. Alice can pay Bob 1
satoshi through Charlie, but cannot do more than 2 satoshis since the payment
channel from Charlie to Bob has a balance of 2 satoshis.}
\label{fig:payment-channel-network}
\end{figure}

\parabf{Offchain networks.} It is clearly impractical for a user to
open a payment channel with every party it needs to transact with; the channel
opening cost and the latency of doing this on the chain would be prohibitive.
Payment channel networks, or {\em offchain networks}, are therefore developed
to support indirect payments between two parties that does not have a payment
channel. An offchain network is composed of many payment channels.
Figure~\ref{fig:payment-channel-network} shows an example of a simple offchain
network with just two payment channels. Two parties can make a transaction so
long as there is a directed path between them and the payment amount is no
bigger than the minimum channel balance of the path. In order to guarantee the
atomicity and security of payments via multiple payment channels, an offchain
network usually relies on the hash time-locked contracts (HTLC)
\cite{poon2016bitcoin}. For example, in
Figure~\ref{fig:payment-channel-network} if Alice wants to pay Bob 1 satoshi 
through Charlie, HTLC guarantees that Charlie receives funds from Alice if and
only if Bob receives the payment from Charlie successfully. Otherwise the
funds are returned to the payer Alice. HTLC also guarantees that either the
balances of all channels on the path are updated or none is updated after the
transaction.

Offchain networks have seen rapid development over recent years and is
increasingly adopted in many scenarios. Lightning Network
\cite{lightningnetwork}, Raiden Network \cite{raiden}, and
Ripple \cite{ripple} are prominent examples in practice. Lightning and Raiden
are offchain networks for Bitcoin and Ethereum, two of the most popular
blockchain technologies, respectively. Lightning for example has 2,700+
active nodes, 21,000+ channels, and 560+ bitcoins ($\sim$\$2M USD)
network capacity as of January 2019 according to \cite{lightning-stats}.
Ripple is another large offchain network using its own cryptocurrency called XRP
as the main transaction medium. Its network has 200+ enterprise customers
including banks and payment providers worldwide. All three offchain networks
allow transactions involving multiple payment channels.

\subsection{Motivation}
\label{sec:motivation}

We believe a sensible first step of designing offchain network routing 
is to understand the workloads carried by these newly emerged networks,
that is the cryptocurrency transactions. Unfortunately this aspect has received 
little attention so far compared to other features of offchain networks such as
the topological characteristics \cite{roos2018settling}.

We conduct a measurement study of transactions in the Ripple network, which to
our knowledge is the only offchain network whose transaction data are publicly
available. We use a dataset from \cite{ripple-trace} that includes over
2.6 million transactions in Ripple from January 2013 to November
2016. Each transaction data entry includes sender, receiver, volume, and time
information.
In addition we crawl
the onchain transactions of Bitcoin as our second dataset.
We believe the characteristics of onchain and offchain transactions are similar
as more onchain transactions are moving to offchain networks for faster
turnaround and lower cost.
We run a full Bitcoin node with Bitcoin Core\footnote{\url{https://bitcoin.org/en/full-node}}
to synchronize all blocks and collect all the transactions (over 103 million)
from its launch in January 2009 to October 2018.
We exclude all the newly generated coins with no
senders.

We now highlight two unique characteristics of cryptocurrency transactions
observed from the traces.

\begin{figure}[t]
    \centering
     \begin{subfigure}{0.48\linewidth}
     \includegraphics[width = \linewidth]{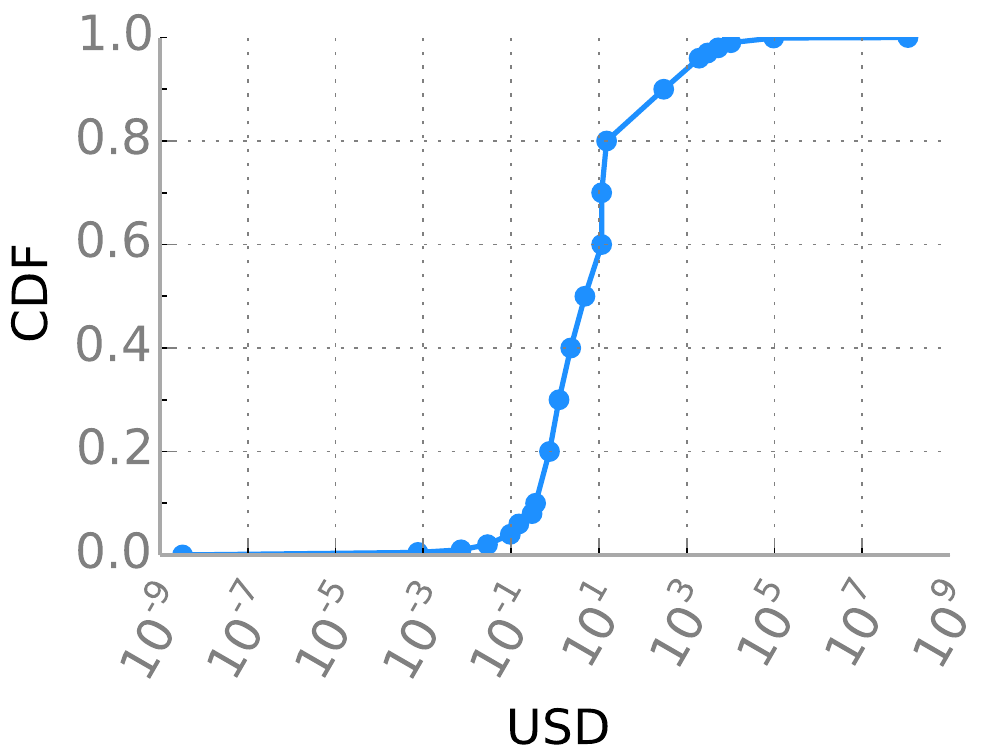}
     \caption{Ripple}
     \label{fig:ripple-size}
     \end{subfigure}
    ~
    \begin{subfigure}{0.48\linewidth}
     \includegraphics[width = \linewidth]{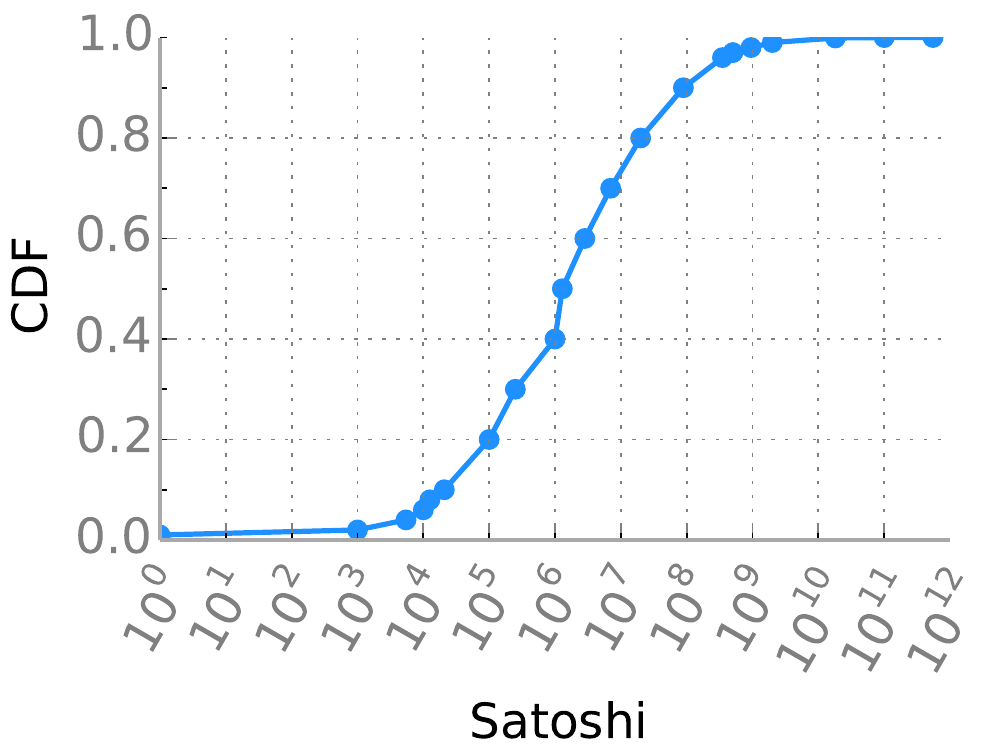}
     \caption{Bitcoin}
     \label{fig:bitcoin-size}
    \end{subfigure}
     \caption{Payment size distributions for Ripple and Bitcoin
     transactions.}
	 \label{fig:payment-size}
\end{figure}%

\parabf{Payment sizes are heavy-tailed.}
We first investigate the payment size distribution.
Figure~\ref{fig:payment-size} shows the CDF of payment size in Ripple and
Bitcoin traces. We observe that most payments are small, while a few
large payments contributes the most volume. For Ripple, nearly 94.5\% of the
volume belongs to only 10\% of the payments, the size of which is more than
\$1,740 USD. The median payment size is only \$4.8 USD.
For Bitcoin, 10\% of payments larger than $8.9\times10^7$ satoshis 
contribute
94.7\% of the total volume, while the median payment size is only
$1.293\times10^6$ satoshis.
This is intuitive to understand since in practice
high-volume transactions happen between a small number of enterprises and
financial institutions. For
example, the volume of transactions between banks can be substantial, but the
trade frequency is relatively low. Most of the time transactions happen
between individuals and merchants, such as money transfer between friends and
families, and purchases of goods and services. These constitute the vast
majority of the use of cryptocurrencies in the same way traditional
currencies are used.

The design implication of heavy-tailed payment size is salient. On one
hand, small or {\em mice payments} are naturally less likely to saturate a
payment channel, and tend to be sensitive to the settlement time. As such,
they can be delivered with high probability using just one or a few paths, and
the paths do not have to be carefully chosen to minimize delay. On the other
hand,
a large payment or {\em elephant payment} consumes much more funds and using a
few paths may not be sufficient. Elephant payments are more
important to the offchain network as their successful delivery would greatly
improve the success volume and performance. At the same time minimizing
transaction fees is also important to elephant payments given their
significant volume. We thus believe a more delicate and optimized routing
solution is justified for elephant payments to thoroughly consider all the
factors above. The solution needs to strategically choose a good set of paths
with enough capacity, and carefully schedule the elephant payment across the
paths with varying fees. The increased settlement time and probing overheads are
acceptable given the low frequency of elephants.

Our distinct treatment of elephant and mice payments is markedly different from
prior work that treats all payments equally through the same routing mechanism
\cite{moreno2017silentwhispers,prihodko2016flare,roos2018settling,sivaraman2018routing}. As we will show, exploiting this
characteristic gives us more flexibility to improve success volume and ratio
of the network while maintaining the overheads.

\begin{figure}[t]
    \centering
        \begin{subfigure}{0.48\linewidth}
     \includegraphics[width = \linewidth]{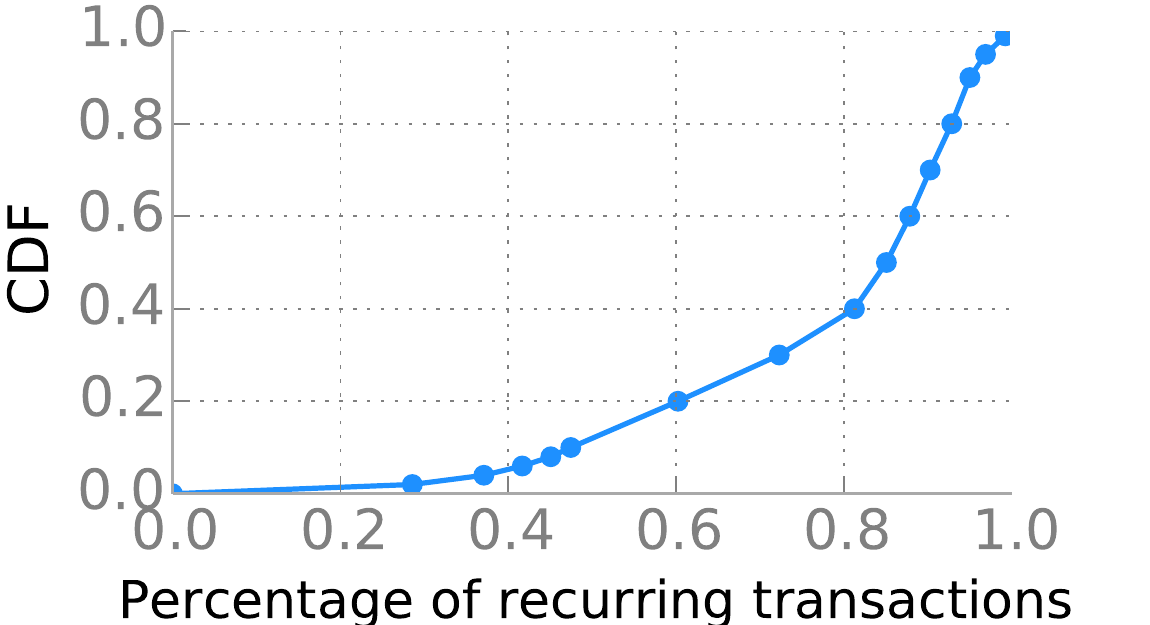}
     \caption{The CDF of percentage of recurring transactions among all
     transactions in a 24-hour period across 1306 days in Ripple.}
     \label{fig:repeated-transactions}
    \end{subfigure}
     ~
\begin{subfigure}{0.48\linewidth}
     \includegraphics[width = \linewidth]{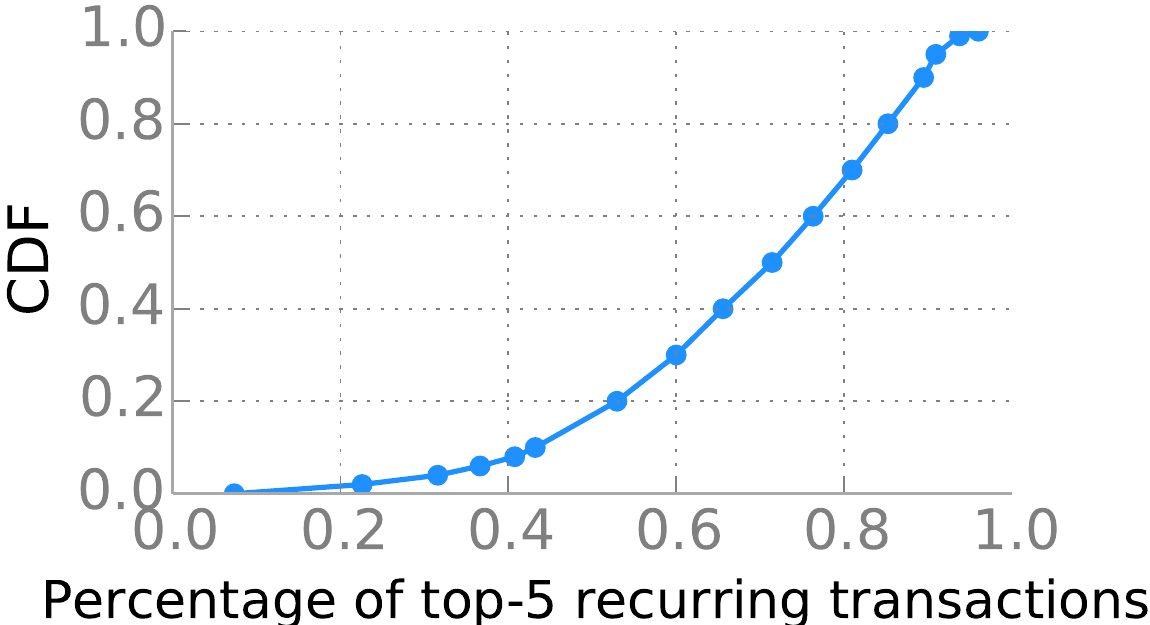}
     \caption{The CDF of percentage of top-5 recurring transactions among
   all recurring transactions in a 24-hour period across 1306 days in Ripple.}
     \label{fig:top-5-repeated-trans}
     \end{subfigure}
\caption{Analysis of the recurring transactions in the Ripple trace.}
\label{fig:recurring-payments}
\end{figure}%

\parabf{Payments are highly recurrent and {clustered}.}
We next investigate the relationship between sender and receiver of the offchain
transactions. Due to the lack of this information in the Bitcoin
trace, we only analyze the Ripple trace.
We examine each of the 1306 days the Ripple trace covers, and identify the
recurring transactions as those with the same sender-receiver pairs within a 
24-hour period. 

Observe that the median percentage of recurring transactions among all
transactions of the day stands at 86\% across 1306 days as shown in
Figure~\ref{fig:repeated-transactions}. Thus most of the transactions in
Ripple are actually recurring within a 24-hour time frame. Moreover, we find
that a user's recurring transactions happen with a small set of users.
Figure~\ref{fig:top-5-repeated-trans} shows for an average user, its top-5
most frequent recurring payments account for over 70\% of the daily
transactions. These properties again make intuitive sense since the real-world
financial relationship for most people is stable and clustered. One mostly
transacts with a small number of parties such as their favorite online
merchants and offline businesses (shops, diners, etc.) near work and home, as
well as their friends and family.

The design implication of recurring transactions and clustered receivers is also
interesting. 
It allows the use of a routing table to store the paths for the recurring
receivers, so path finding can be simplified to table lookups especially for
mice payments. 
A small routing table would be enough to cover most recurring transactions due
to the clustered nature of them. 
This is instrumental towards reducing the overhead of processing (mice) payments
without much performance sacrifice.

% summary
To quickly recap, the transaction characteristics presented here enable us to
explore a larger design space for offchain routing, and motivate our design of
\sys which we now introduce.

\section{Design}
\label{sec:design}

\sys is a distributed online routing system that processes each transaction as
it arrives
at the sender,
because a centralized offline approach is inherently infeasible for
decentralized
offchain networks with constantly changing channel balance.
\sys differentiates elephant and mice payments and applies different routing
algorithms in order to achieve a better performance-overhead tradeoff. For
elephants that have significant impact on overall performance, \sys first adopts
a
modified max-flow algorithm to find and probe paths with sufficient balance
to
satisfy their demands, and then solves an optimization program to split the
payment over paths to minimize the transaction fees. For mice payments whose
demands are easy to satisfy, \sys uses a lightweight solution that simply
routes them randomly through a small set of precomputed paths whenever possible
in order to reduce the
probing overhead.

\subsection{Prerequisites}
\label{sec:model}

\sys's design relies on two prerequisites about the offchain networks.

\parabf{Locally available topology.}
The topology of an offchain network, without the channel balance information,
is fairly stable and changes on an hourly or daily scale. This is because
opening or closing a payment channel requires onchain transactions which take
at least tens of minutes, and a channel usually remains in the network after
establishment. Therefore practical offchain routing protocols in Lightning and
Raiden
require each node to locally store the topology of the offchain network and
periodically update it using some gossiping protocols
\cite{lightingnetworkcode, raidennetworkcode}.
\sys assumes similar mechanisms are in place and the connectivity topology is
locally available at each node. Note the topology is a directed
graph since payment channels are bidirectional: funds can flow in either
direction and channel balances on different directions are different.

\parabf{Atomic multipath payments.}
To improve the network utilization, \sys uses multipath routing whenever
possible and assumes the atomicity of multipath payments is guaranteed,
similar to prior work \cite{sivaraman2018routing}. This can be achieved by
mechanisms such as Atomic Multipath Payments (AMP) proposed for Lightning
\cite{amp}. Building upon HTLC, AMP allows a payment to be split over multiple
paths while ensuring the receiver either receives all funds from several
partial payments, or gets nothing (i.e. payment fails). The design and
implementation of such a mechanism is beyond this paper.

\subsection{Routing Elephant Payments}
\label{sec:elephants}

The design challenges for routing elephant payments are: (1) how to find good
paths with sufficient capacity to satisfy demand as much as possible, and (2)
how to carefully split the payment
across the paths in order to minimize the transactions fees.

\parabf{Path finding with modified max-flow.}
We discuss some strawman solutions to the first challenge on path finding and
why they do not work, and then present \sys's solution with a modified max-flow
algorithm.

\begin{figure}[t]
\centerline{\includegraphics[width=0.48\textwidth]{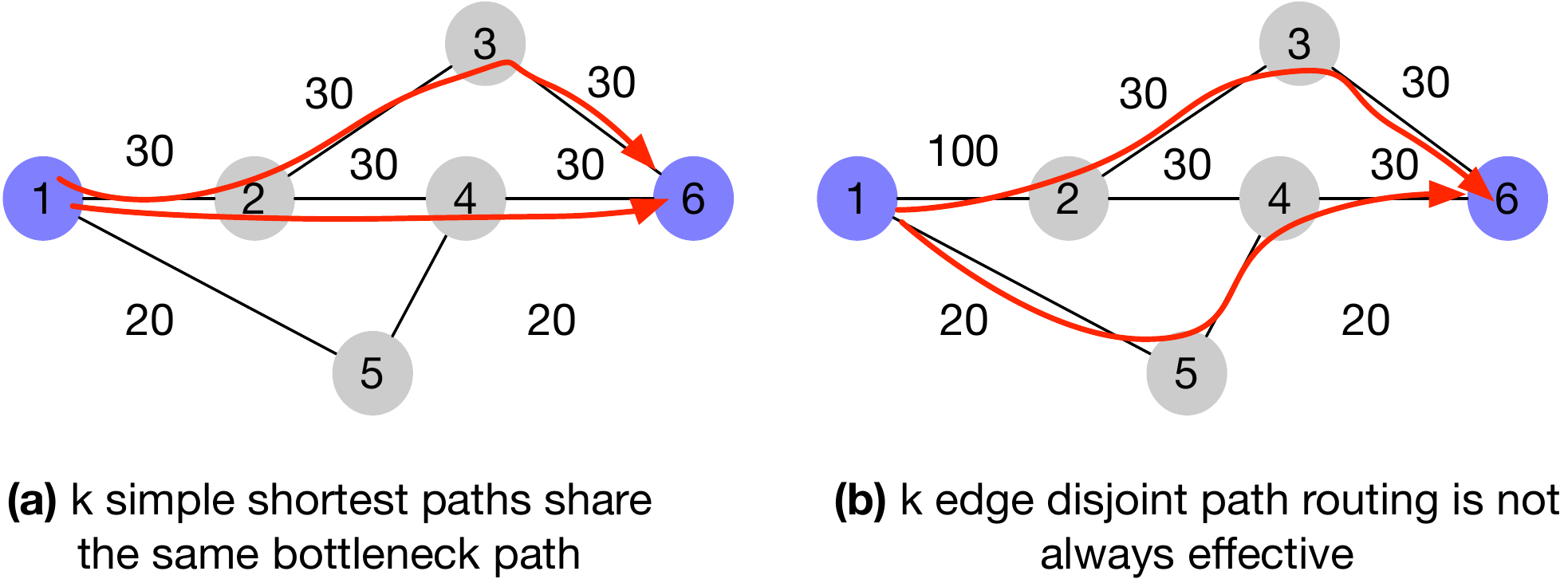}}
\caption{An illustrative example of common shortest path schemes. Node 1 is the sender, and node 6 is the receiver. In each scheme two
paths are used, i.e. $k=2$.}
\label{fig:shortest-path}
\end{figure}

\parait{Strawman solutions.}
With the network topology locally available, a first attempt at the path
finding problem would be to simply have the sender compute $k$ good paths.
Shortest paths for example are a natural choice since they minimize number of
hops and helps reduce transaction fees. However, restricting to shortest paths
may lead to severe underutilization when they share a common bottleneck. To
see this, we consider an example in Figure~\ref{fig:shortest-path}(a). Two simple
shortest paths from node 1 to 6 share the same bottleneck link from node 1 to
2. Using them provides a total capacity of 30 while the other path of 1-5-4-6
is underutilized. To overcome this one may consider edge-disjoint shortest
paths, which are used in Spider \cite{sivaraman2018routing}. Yet they may not
always work either especially when the common bottleneck has abundant capacity.
Figure~\ref{fig:shortest-path}(b) shows that using 2 edge-disjoint shortest
paths yields a total capacity of 50, while using 2 simple shortest paths that
traverse from 1 to 2 yields a total capacity of 60 since the common link from
1 to 2 has abundant capacity now.

It is thus important to consider channel capacity in path finding for elephant
payments. This naturally motivates us to resort to max-flow algorithms
\cite{ford1956maximal}. A max-flow algorithm such as Edmonds-Karp
\cite{CLRS09} is used to find the maximum flow between a pair of
nodes in a flow network. However they cannot be directly applied to offchain
networks. Max-flow algorithms require a weighted graph, meaning that the
balance or capacity of all edges of the graph should be known. This is
infeasible in our problem: the channel balance is dynamically changing in
offchain networks, and probing each channel of each path whenever an elephant
payment arrives does not scale for a network with thousands of nodes and tens
of thousands of channels \cite{lightning-stats}.

\parait{\sys's solution.}
We thus develop a modified max-flow algorithm based on Edmonds-Karp \cite{CLRS09} to
sequentially find $k$ paths and their maximum flow without excessive
overheads. Algorithm~\ref{alg:pathfinding} shows the pseudocode.

\alglanguage{pseudocode}
\begin{algorithm}[t]
\small
\caption{Modified Edmonds-Karp for elephant payment routing}
\label{alg:pathfinding}
\begin{algorithmic}[1]
\State {{\bf Input}: Topology graph $G$, a payment ($s$, $t$, $d$) from $s$ to
$t$
with demand $d$, maximum number of paths needed $k$}
\State{{\bf Output}: Path set $P$, capacity matrix $C$}

\State {$P =\emptyset, f=0$  \Comment{Initialize maximum flow $f$} }
\State {$C = \infty$ \Comment{Initialize capacity matrix $C[n \times n]$ }}
\State {$C' = \infty$ \Comment{Initialize residual capacity matrix $C'[n \times
n]$ }}
\While {$|P|<k$} \Comment {Find at most $k$ paths}
  \State {$p = $ Breath-First-Search$(G, C', s, t)$}
  \If {$p == \emptyset$}
    \State {\bf break}
  \EndIf
  \State {Add $p$ to $P$}
  \State {Probe each channel on $p$ to obtain their capacity $C_p$}
  \State {Find the bottleneck capacity $c=\min\ C_p$}
  \State {$f = f+ c$}
  \State {$v = t$}
  \For {$v\neq s$}
    \State {$u = p[v]$}
    \If { $C[u,v] = \infty$ } \Comment{Set channel capacity for the
      first time}
      \State {$C[u,v] = C_p[u,v]$}
      \State {$C'[u,v] = C_p[u,v]$}
    \EndIf
    \If { $C[v,u] = \infty$ }
      \State {$C[v,u] = C_p[v,u]$}
      \State {$C'[v,u] = C_p[v,u]$}
    \EndIf
    \State $C'(u, v) = C'(u, v)-c$ \Comment{Reduce channel capacity }
    \State $C'(v, u) = C'(v, u)+c$ \Comment{Increase capacity of the channel in
    the reverse direction}
  \EndFor
\EndWhile
\If {$f \ge d$}
  \State\Return $P, C$ \Comment{Return paths found and capacity}
\Else
  \State \Return $\emptyset$
  \EndIf
\Statex
\end{algorithmic}
\end{algorithm}

Each node has the network topology $G$ without capacity information.
When a new elephant arrives the sender $s$ invokes Algorithm~
\ref{alg:pathfinding} to route it.
It uses a capacity matrix $C$ to record
the probed channel capacity of the paths, and a residual capacity matrix $C'$
to record the remaining capacity of channels as in Edmonds-Karp
\cite{CLRS09}. Both $C$ and $C'$ are initialized to infinity 
(lines 4--5 in Algorithm~\ref{alg:pathfinding}).
It then enters a loop with at most $k$ iterations to find at most $k$ paths. In
each loop \sys first runs the Breadth-First-Search on topology $G$ with
the residual capacity matrix $C'$ to find a
feasible shortest path $p$ with non-zero capacity (line 8), and adds $p$ to the solution set $P$.
It then sends probes along $p$ to obtain capacity of each channel on it, and
obtains the bottleneck capacity $c$. This indicates that we can send $c$ on path
$p$ (line 14).
It updates the capacity of channels that have been probed for the first time in
$C$ according to the probing results $C_p$.
It also updates the residual capacity of channels on path $p$ in the residual
capacity matrix $C'$ using
$c$ to reflect the new flow found by $p$.
After the loop terminates, Algorithm~\ref{alg:pathfinding} returns the paths $P$
and capacity matrix $C$ if the maximum flow $f$ over these paths satisfies the
payment demand $d$.

Compared to Edmonds-Karp with $O(|V||E|)$ iterations, our algorithm ends with
at most $k$ iterations and $k$ paths to probe when there are at least $k$ paths
between $s$ and $t$ on $G$.
This helps reduce the probing overhead.
We find that setting $k$ between 20 to 30 provides good performance in practical
offchain network topologies with thousands of nodes and tens of thousands of
channels.
Also our algorithm works without the capacity matrix as input by
assuming each channel has non-zero capacity. It is thus possible, though rare
in our evaluation, that our algorithm finds a path but its effective capacity is
zero after probing.

\parabf{Path selection.}
Given a set of paths with sufficient capacity from Algorithm~\ref{alg:pathfinding}, 
the next step is to determine how to route over them
to minimize the total transaction fees.
The fee information is collected during the
probing process with the capacity information.
We take a principled approach and solve this using mathematical optimization.

Specifically, we have the path set $P$ and the capacity matrix $C$. We
represent the fee collected by a channel $(u,v)$ with a charging function
$f_{u,v}$. We assume $f$ is convex. Thus the fee amounts to $f_{u,v}(r_p)$ if we
route a partial payment of $r_p$ to $(u,v)$. The objective of the optimization
program is to minimize the total fees subject to constraints that the payment
demand $d$ is met, and channel capacity is respected:
\begin{align}
\min \quad
& {\sum_{p \in P} \sum_{\{(u, v)\}}a^p_{u, v}f_{u, v}{(r_p)}} \label{opt:cost}
\\
\text{subject to}\quad
&\sum_{p \in P}{r_p} = d, \nonumber\\
&\sum_{p \in P}{r_p a^p_{u, v}} - \sum_{p \in P}{r_p a^p_
{v, u}} \leqslant {C(u, v)}, \forall (u,v). \nonumber
% & {-c(v, u)}  \leqslant .
\end{align}
Here $a^p_{u, v}$ indicates whether $p$ uses channel $(u,v)$ or not. 
Note that partial payments on different direction of the same channel can
offset each other in terms of balance.

The optimization program \eqref{opt:cost} is a convex optimization and can be
solved using standard solvers quickly due to the small problem size with 
$k$ paths. In practice the fee charging function is typically linear with a
fixed fee plus a volume-dependent component, which means \eqref{opt:cost} is a
simple linear program and even easier to solve.

\subsection{Routing Mice Payments}
\label{sec:mice}

The design challenge for mice payment routing is to simplify the protocol and
minimize overhead due to their large quantity.
Applying elephant routing design here would be an overkill.
We now present our design for mice payments which also consists of path finding and path selection.

\parabf{Path finding.}
Each node maintains a routing table for mice payments. It contains paths for
the unique receivers of this node. Upon seeing a new receiver that does not
exist in the routing table, the node computes top-$m$ shortest paths (i.e.
using Yen's algorithm \cite{yen1971finding}) on the local topology $G$, and
adds them to the routing table. If the receiver is in the routing table, \sys
simply re-uses the existing paths. Since most payments are recurring as
explained in \cref{sec:motivation}, this design simplifies path finding into
table lookups in most cases without any computation. The recurring
nature of mice also ensures the routing table size is not too large. We use
top-$m$ shortest paths where $m$ is much less than $k$ the number of paths
used for elephant routing in \cref{sec:elephants}, because mice payments do not
require much capacity, and typically a few shortest paths provide good
performance ($m=4$ in our evaluation).

The routing table is periodically refreshed when the local network topology
$G$ is updated (by the underlying gossip protocol): all entries are re-computed
using the latest $G$. Also when a
payment encounters an unaccessible path with zero effective capacity or no
connectivity, \sys replaces it with the next top shortest path. Timeouts are
used to remove receivers and their
entries that have not been accessed for a long time to limit the routing table size.

\parabf{Path selection.}
With $m$ shortest paths from the routing table, the sender determines path
selection using a
trial-and-error loop. It first sends the full payment along a random path
$p$. If successful the protocol ends. Otherwise, the sender probes $p$ to find
its
effective capacity $c_p$ and sends a partial payment of volume $c_p$ along
$p$. It then updates the remaining demand of the payment and continues the
iteration. 
This ensures low probing overhead since \sys only probes when
it is necessary and at most $m$ paths are probed. The use of multiple paths also
improves the success ratio of
delivering the payment. Instead of following a fixed order (say in a
descending order of path length), \sys randomly picks the paths to better load
balance them without knowing their instantaneous capacities. Lastly, when
$m$ paths are exhausted and demand is not satisfied, \sys declares the payment
fails.

\section{Simulation}
\label{sec:simulation}

In this section, we evaluate the performance of \sys against existing offchain
routing algorithms using simulation. Our evaluation aims to answer the following
questions:

\begin{compactitem}
\item How does \sys perform under realistic offchain network topologies and
traces?
\item How does channel capacity and network load affect \sys's performance?
\item How effective is differentiating elephant and mice payments in \sys?
\item How effective is the mice payment routing algorithm?
\end{compactitem}

\begin{figure*}[ht]
\vspace{-0.3cm}
     \centering
      \begin{subfigure}[t]{0.24\linewidth}
         \centering
         \includegraphics[width=\textwidth]{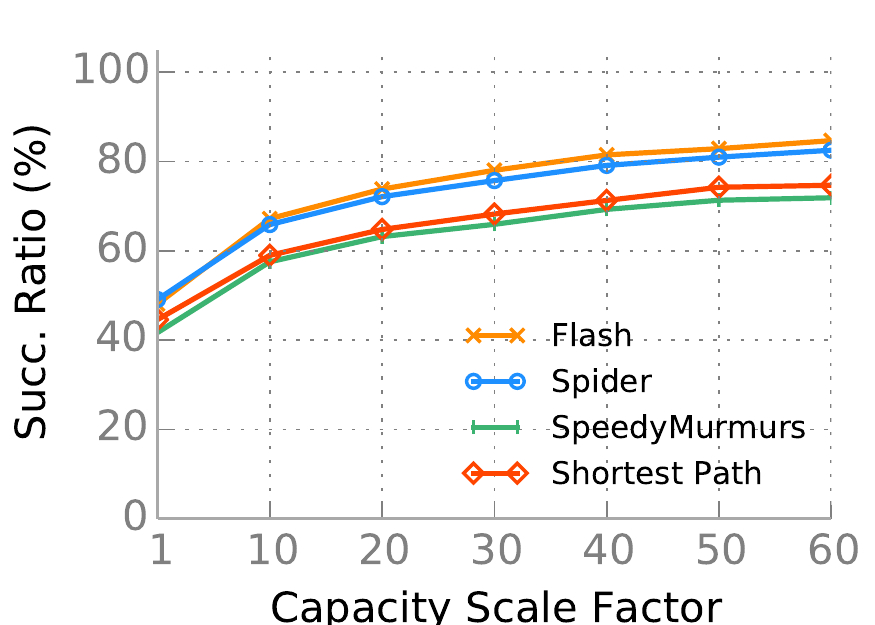}
         \caption{Succ. ratio in Ripple }
     \end{subfigure}
     ~
     \begin{subfigure}[t]{0.24\linewidth}
         \centering
         \includegraphics[width=\textwidth]{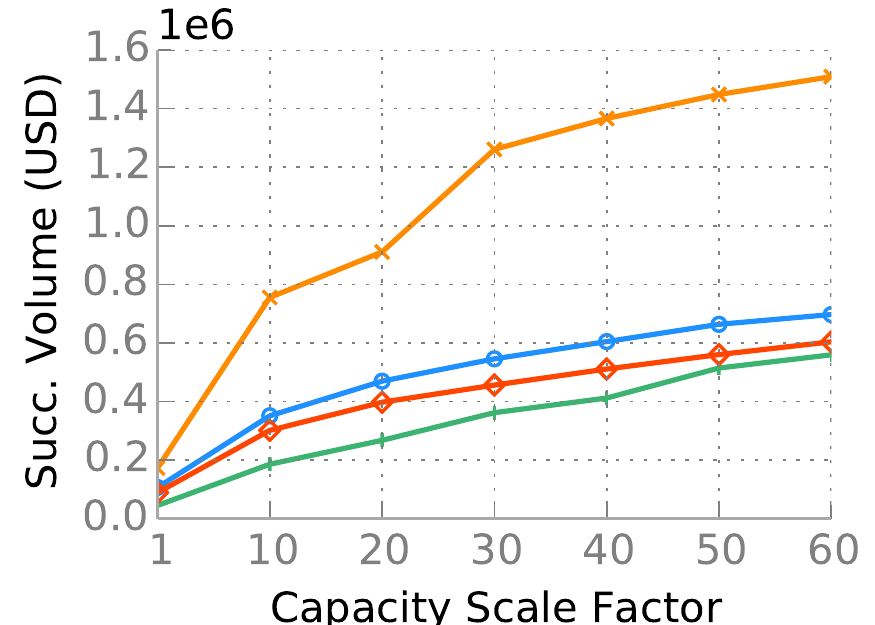}
         \caption{Succ. volume in Ripple }
     \end{subfigure}%
     ~
      \begin{subfigure}[t]{0.24\linewidth}
         \centering
         \includegraphics[width=\textwidth]{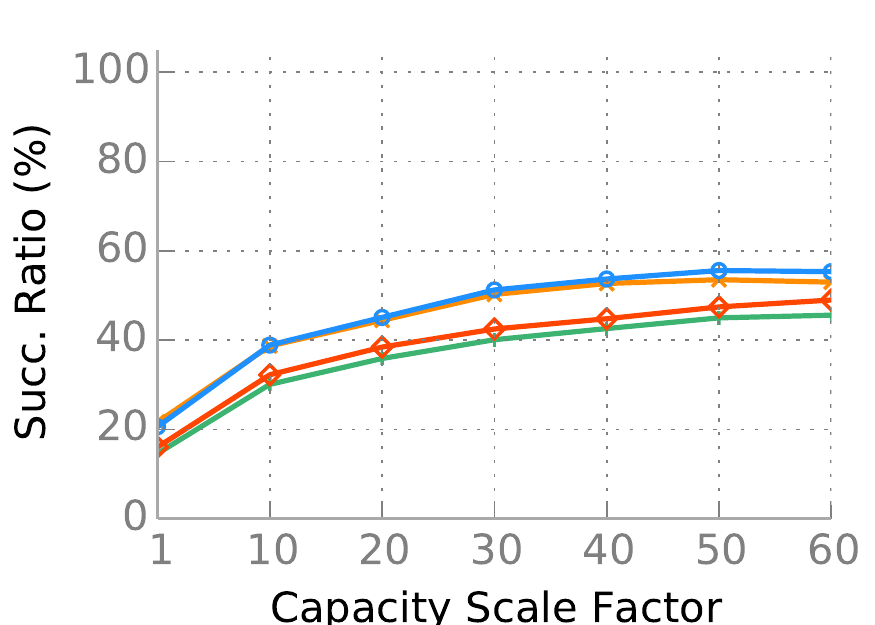}
         \caption{Succ. ratio in Lightning }
     \end{subfigure}
     ~
     \begin{subfigure}[t]{0.24\linewidth}
         \centering
         \includegraphics[width=\textwidth]{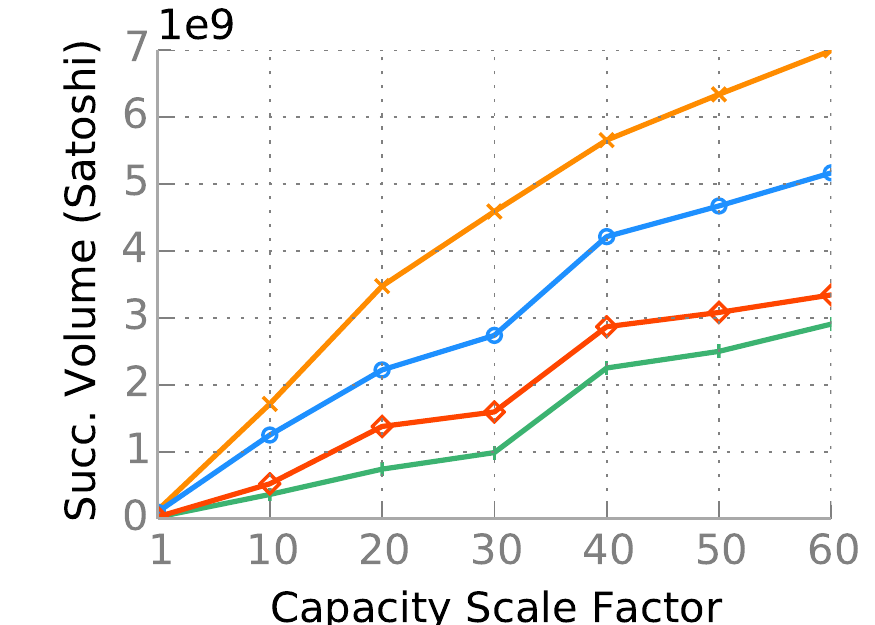}
         \caption{Succ. volume in Lightning}
     \end{subfigure}%
     \vspace{-0.2cm}
     \caption{Performance results with varying link capacities in Ripple and
     Lightning.}
     \label{fig:general-scale}
 \end{figure*}

 \begin{figure*}[t]
\vspace{-0.3cm}
     \centering
     \begin{subfigure}[t]{0.24\linewidth}
         \centering
         \includegraphics[width=\textwidth]{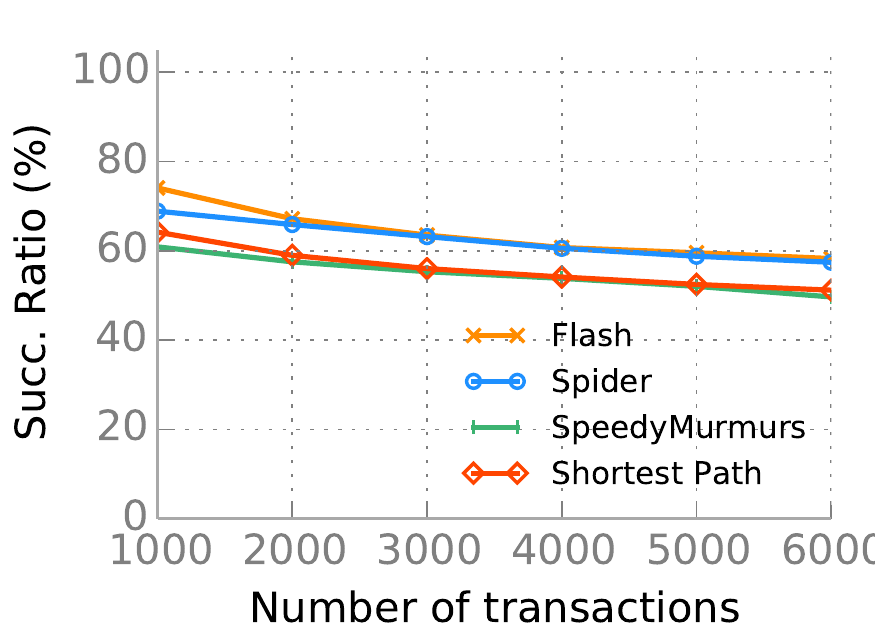}
         \caption{Succ. ratio in Ripple }
     \end{subfigure}
     ~
     \begin{subfigure}[t]{0.24\linewidth}
         \centering
         \includegraphics[width=\textwidth]{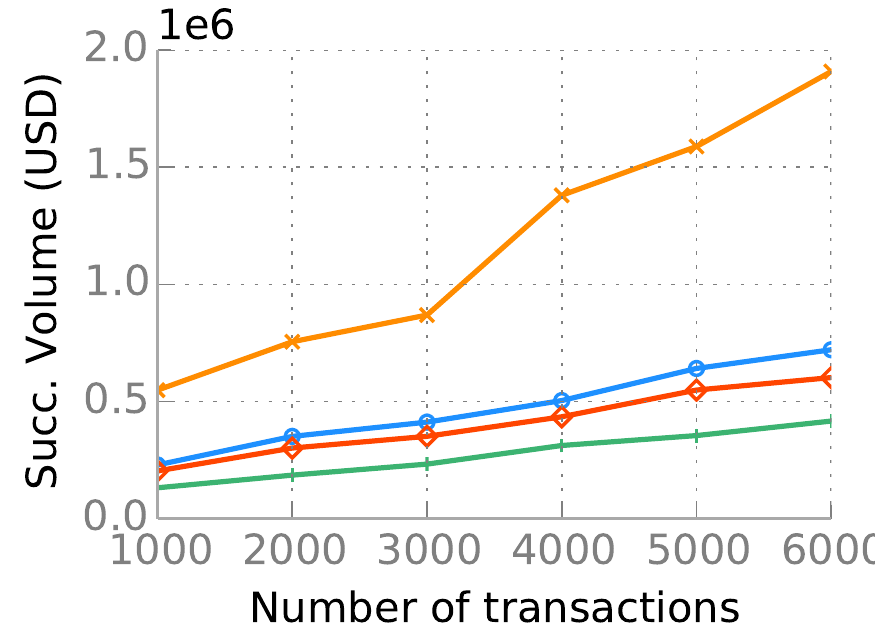}
         \caption{Succ. volume in Ripple }
     \end{subfigure}%
     ~
    \begin{subfigure}[t]{0.24\linewidth}
     \centering
     \includegraphics[width=\textwidth]{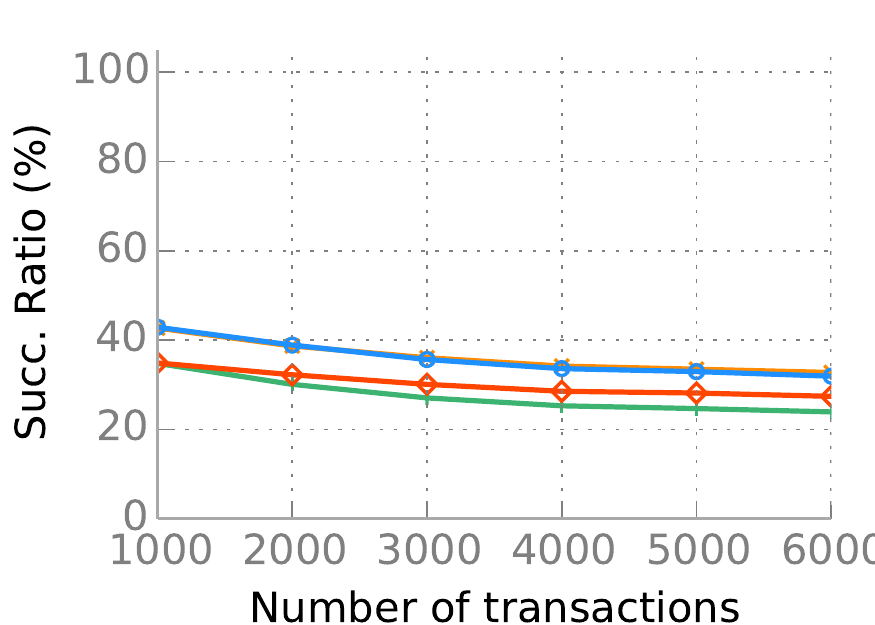}
     \caption{Succ. ratio in Lightning }
     \end{subfigure}
     ~
     \begin{subfigure}[t]{0.24\linewidth}
         \centering
         \includegraphics[width=\textwidth]{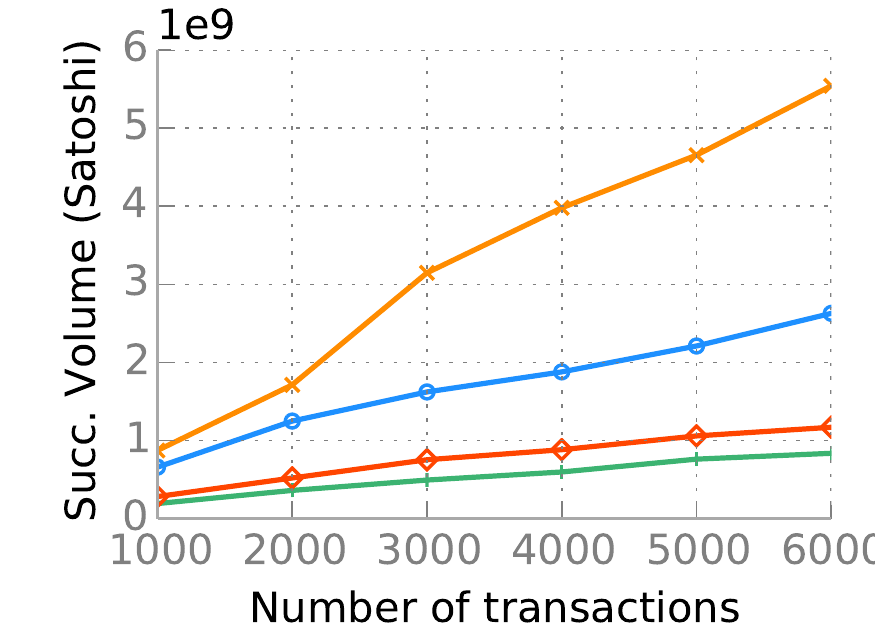}
         \caption{Succ. volume in Lightning }
     \end{subfigure}%
     \vspace{-0.2cm}
     \caption{Performance results with varying number of transactions in Ripple
     and Lightning.} 
     \label{fig:general-load}
 \end{figure*}

\subsection{Methodology}
\label{sec:sim_setup}

\parabf{Setup.}
We implement offchain network topologies and routing schemes using the NetworkX
package in Python \cite{networkx} in the simulation. Our simulation focuses on
evaluating the routing performance in a large-scale real offchain
network, and does not concern the implementation of the underlying security
mechanism (say HTLC).

We evaluate \sys with two real-world offchain network topologies: 
Ripple and Lightning. 
We obtain crawls of Ripple's active nodes and channels from January 2013 to
November 2016 from \cite{ripple-trace}. This topology includes 93,502 nodes and
331,096 edges. We removes nodes with only a single neighbor and channels with no
funds from the topology. The processed topology we use in the simulation
includes 1,870 nodes and 17,416 edges. The distribution of funds on
payment channels in Ripple is extremely skewed. Thus we redistribute the funds
by evenly assigning the total funds over both directions of a channel.
For Lightning topology we run the {\tt c-lightning} \cite{clightning} node on
{\tt mainnet} and connect it to an existing node by opening a channel with the
node. We use commands {\tt listchannels} and {\tt listnodes} to get 
information of nodes and channels as a snapshot of the Lightning network on a
particular day of December 2018. The number of nodes is 2,511 and number of
channels is 36,016. We use the crawled distribution of funds on channels directly.

We generate payments by randomly sampling the Ripple trace for the Ripple
topology. 
Due to the lack of sender-receiver information in the Bitcoin trace for
Lightning, we
randomly sample the Bitcoin trace for transaction volumes, and sample a sender- 
receiver pair from the Ripple trace and map it to nodes in the Lightning
topology. Payments arrive at senders sequentially.

\parabf{Benchmarks.} We compare four offchain routing algorithms.
\begin{compactitem}
\item {\em \sys}: Our routing algorithms. Unless stated otherwise, we set the
number of shortest paths for each receiver in mice payment routing to 4, i.e.
$m=4$, and the number of paths for elephant routing to 20, i.e. $k=20$. The
elephant-mice threshold is set such that 90\% of payments are mice.
\item {\em Spider} \cite{sivaraman2018routing}: The state-of-the-art
offchain routing algorithm which considers
the dynamics of channel balance. It balances paths by using those 
with maximum available capacity, following a ``waterfilling'' heuristic. It
uses 4 edge-disjoint paths for each payment.
\item {\em SpeedyMurmurs} \cite{roos2018settling}: An embedding-based routing
algorithm which relies on assigning coordinates to nodes to find short paths
with reduced overhead.
The number of landmarks is 3 as \cite{roos2018settling} suggests.
\item {\em Shortest Path (SP)}: This is the baseline. SP uses the path with
the fewest hops between the sender and receiver to route a payment.
\end{compactitem}

\parabf{Metrics.} Similar to prior work \cite{roos2018settling,
sivaraman2018routing}, we use success ratio, success volume and number of
probing messages as the primary metrics in the simulation. We report the average
results
over 5 runs.

\subsection{Overall Performance and Overhead}
\label{sec:overall}
We now examine the performance and overhead of \sys with different settings of
the offchain network.

\parabf{Performance with different capacities.} 
We first evaluate the
performance of \sys with various link capacities. The medium channel capacity in
Lightning is around 500,000 Satoshi and in Ripple is 250 USD. As offchain
networks are still in their infancy and the capacity provided may be limited, we
scale the link capacity by a factor of 1 to 60 in the simulation similar to
existing work \cite{roos2018settling,sivaraman2018routing}. 
The number of transactions used is fixed at 2000. 
Figure~\ref{fig:general-scale} shows the success ratio and volume results. 
For both Ripple and Lightning, \sys performs $\sim$20\% better than
SpeedyMurmurs and Shortest Path on success ratio. 
\sys and Spider are both able to fulfill most mice payments. 
As the success ratio is dominated by mice payments, \sys and Spider achieve similar performance. 
For success volume, \sys performs up to 4.5x better over Shortest Path,
5x better than SpeedyMurmurs, and 2.3x better than Spider. 
The success volume benefits of \sys is due to its delicate elephant payment
routing that uses more capacities and carefully schedule the partial payments to
deliver them successfully. 
As the network capacity increases, we
observe more successful payments and thus the increase of both success ratio and volume. \sys consistently
outperforms other schemes. 

\parabf{Performance with varying transaction numbers.} 
We also vary the number
of transactions flowing into the network to emulate different loads. 
The capacity scale factor is 10. 
With the increase of number of transactions, the success ratio of all
schemes degrades as shown in Figure~\ref{fig:general-load}. 
One possible reason is that, as more
payments especially elephant payments are accepted, channels are easier to be saturated in one direction. Although the number of successful payments keep increasing, the probability to fulfill a payment decreases. 
Observe that \sys consistently outperforms other schemes. It shows
significant benefits on success volume: the performance gains over Shortest
Path, SpeedyMurmurs, and Spider are up to 4.7x, 6.6x, and 2.6x, respectively.
We also observe that \sys's performance gains increase with more
transactions, suggesting that it scales better than other solutions. 

\begin{figure}[t]
    \vspace{-0.3cm}
     \centering
      \begin{subfigure}[t]{0.48\linewidth}
         \centering
         \includegraphics[width=\textwidth]{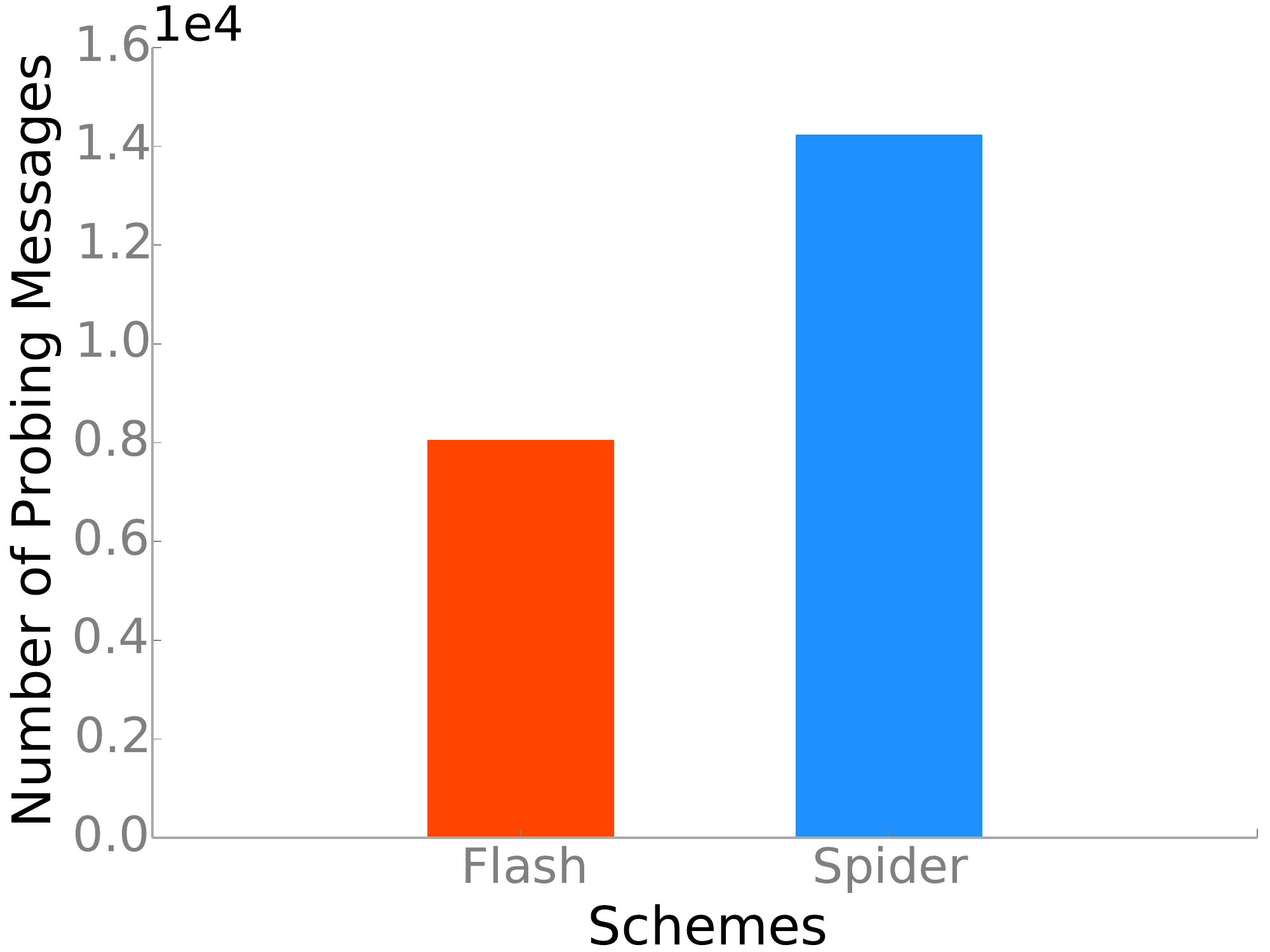}
         \caption{Ripple}
     \end{subfigure}
     ~
     \begin{subfigure}[t]{0.48\linewidth}
         \centering
         \includegraphics[width=\textwidth]{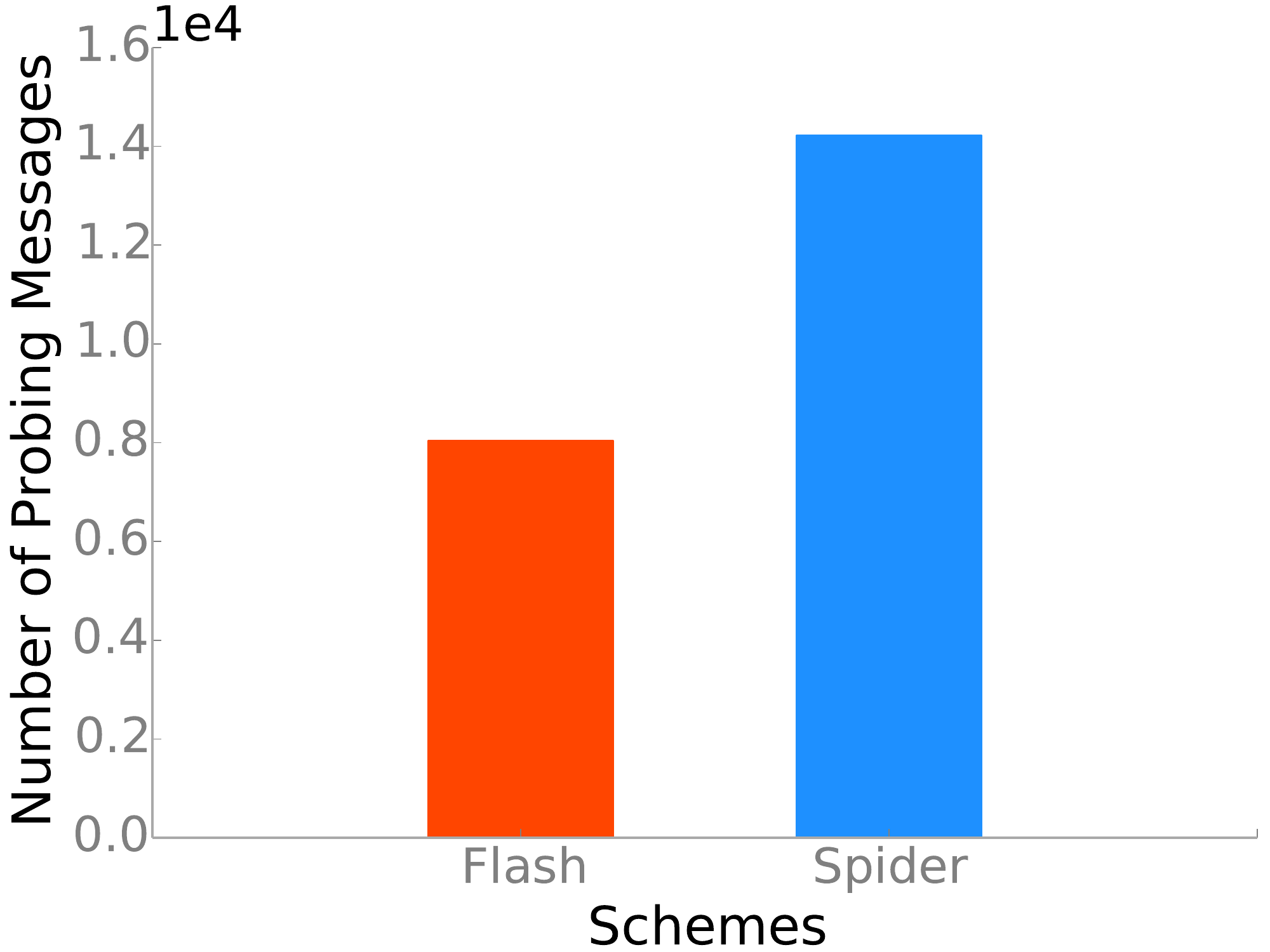}
         \caption{Lightning}
     \end{subfigure}
     ~
     \vspace{-0.2cm}
     \caption{Probing message comparison results.}
     \label{fig:general-overhead}
 \end{figure}

\parabf{Probing message overhead.} We have demonstrated the performance
improvement of \sys in terms of success ratio and volume. 
We now evaluate the number of probing messages of \sys to see if our
algorithms can curb the overhead of routing. 
Figure~\ref{fig:general-overhead} shows the comparison results with 2000
transactions and a capacity scale factor of 10.
Note that SpeedyMurmurs and Shortest Path are static routing schemes without
probing.  
Without probing they suffer from poor performance as discussed just now. 
We thus exclude them from the comparison here. 
The number of probing messages along a path is proportional to the number
of hops of the path.

Observing from Figure~\ref{fig:general-overhead}, compared to Spider which also
uses multiple paths, \sys saves 43\% message overhead in Ripple and 37\% in Lightning. 
Spider treats mice and elephant flows the same and always uses 4 shortest paths.
\sys differentiates mice and elephants: though it uses much more paths (20) for
elephants, it uses at most 4 paths for the vast majority of the mice payments
in order to balance the performance and overhead tradeoff. 
Moreover, \sys's mice payment routing relies on a trial-and-error approach to
further
reduce probing overhead: it only probes a path when it cannot deliver the
payment in full, which usually do not happen for mice payments. 
We observe that most mice payments are delivered with 1 or 2 paths.
Thus the results here demonstrate that \sys indeed achieves a better tradeoff
between performance and overhead compared to state of the art.

  \begin{figure}[t]
	\vspace{-0.3cm}
     \centering
      \begin{subfigure}[t]{0.485\linewidth}
         \centering
         \includegraphics[width=\textwidth]{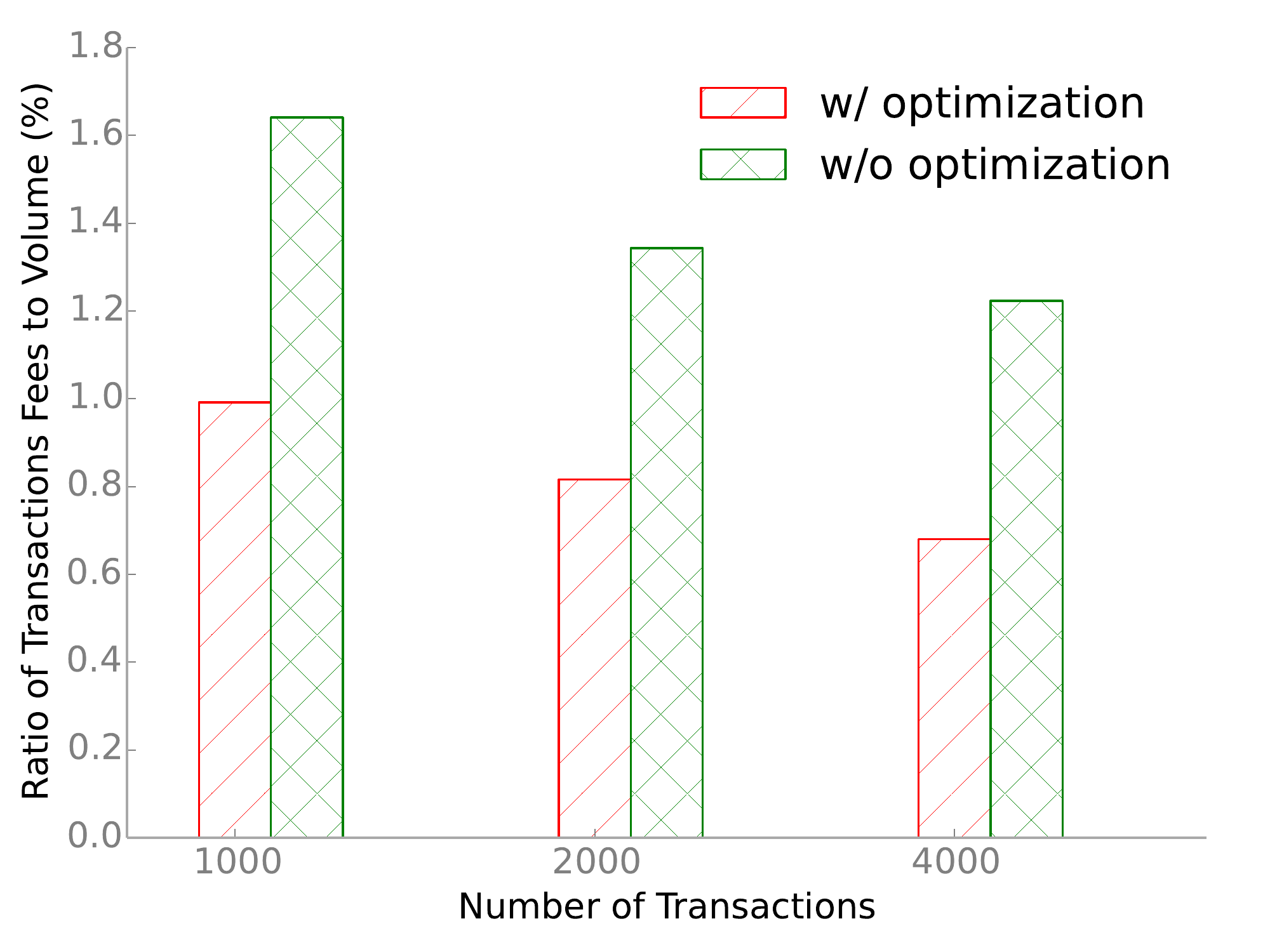}
         \caption{Lightning}
     \end{subfigure}
     ~
     \begin{subfigure}[t]{0.485\linewidth}
         \centering
         \includegraphics[width=\textwidth]{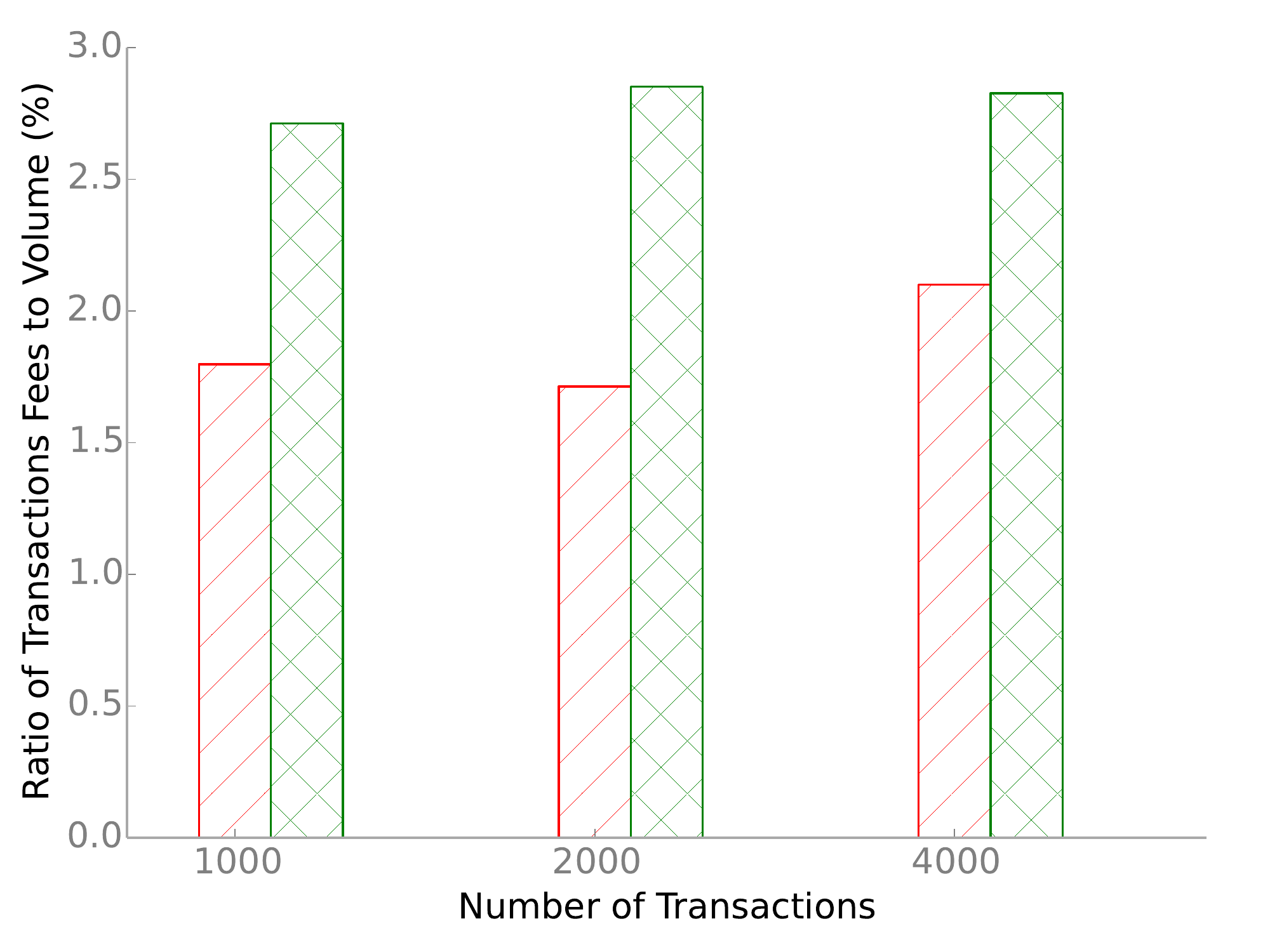}
         \caption{Ripple}
     \end{subfigure}
     \vspace{-0.2cm}
     \caption{Impact of transaction fee optimization in \sys. }
     \label{fig:micro-cost}
 \end{figure}

\subsection{\sys Microbenchmarks}

We now take a deep dive into \sys by evaluating microbenchmarks about the impact
of some key parameters to its design. 
Through the microbenchmarks, we also verify our design choices. In all
experiments here we use 2000 transactions in each run and a capacity scale
factor of 10 unless stated otherwise.

\parabf{Impact of transaction fee optimization.}
As mentioned in \cref{sec:elephants}, \sys splits an elephant payment over
multiple paths to minimize the total transaction fees. We now validate the
effectiveness of this design. 
To perform a fair comparison, we realize \sys
without transaction fees minimization as the baseline, where the paths are used
sequentially as they are found by our modified Edmonds-Karp algorithm until the
demand is met.
We compare the unit transaction fees (in percentage) to
avoid the impact of volume on the result. Note the unit fee is obtained over all
payments, not just elephant payments. 
We set 90\% channels with a random fees from 0.1\% to 1\% and
10\% channels from 1\% to 10\% of the volume. 
Observe from Figure~\ref{fig:micro-cost} that \sys reduces the transaction cost
by around 40\% on average in both Ripple and Lighting compared to not performing
fee minimization. 

\begin{figure}[t]
    \vspace{-0.3cm}
     \centering
      \begin{subfigure}[t]{0.485\linewidth}
         \centering
         \includegraphics[width=\textwidth]{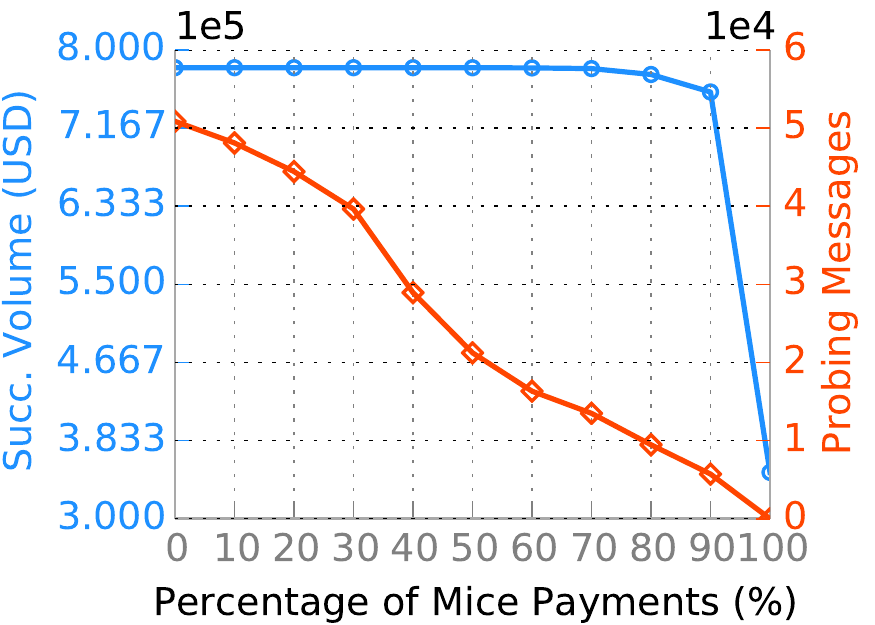}
         \caption{Ripple }
     \end{subfigure}
     ~
     \begin{subfigure}[t]{0.485\linewidth}
         \centering
         \includegraphics[width=\textwidth]{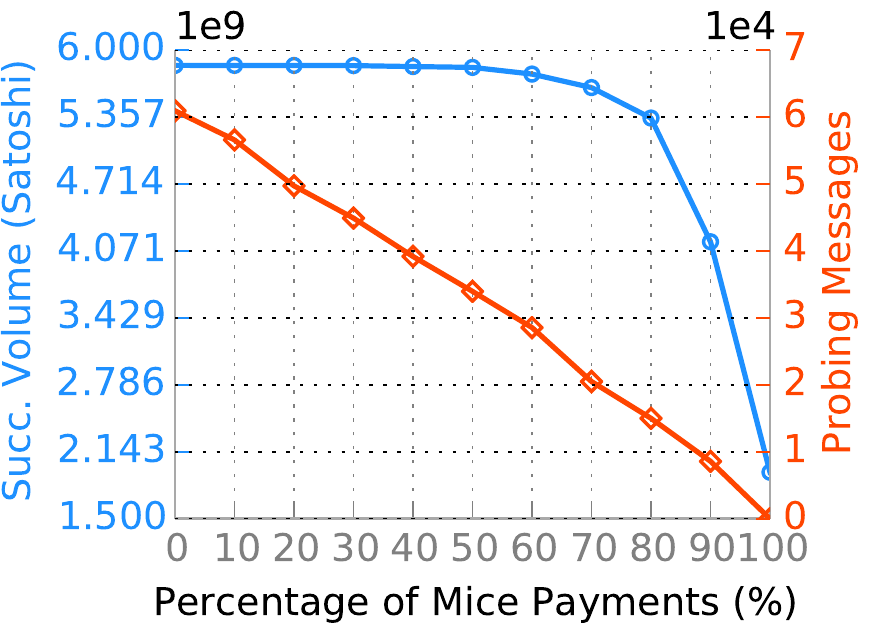}
         \caption{Lightning}
     \end{subfigure}
     \vspace{-0.2cm}
     \caption{Impact of threshold value in \sys. }
     \label{fig:ripple-threshold}
 \end{figure}

\parabf{Impact of threshold.}
We first show how the choice of threshold impacts the performance, i.e. success
volume of payments. 
Here we vary the threshold value such that the percentage of mice payments
varies from 0\% to 100\%.
Obviously a higher percentage with a larger threshold results in more payments
classified as mice. 
Observe from Figure~\ref{fig:ripple-threshold} that the success volume of mice
payments remains stable until the percentage of mice reaches 80\%--90\%.
That is, when most payments are treated as mice with \sys's simple routing
algorithm, their success volume is only marginally smaller than when everyone is
treated by the elephant routing algorithm. 
However, the probing overhead increases as the percentage of mice
payments decreases and probing is more aggressively used. 
This clearly demonstrates that our design choice of differentiating mice and
elephant is effective: it significantly reduces the probing overhead without
much performance degradation for most mice payments. 
This also justifies our setting of threshold with 90\% mice flows which achieves
a good performance-overhead tradeoff.

\begin{figure}[t]
\vspace{-0.3cm}
     \centering
      \begin{subfigure}[t]{0.485\linewidth}
         \centering
         \includegraphics[width=\textwidth]{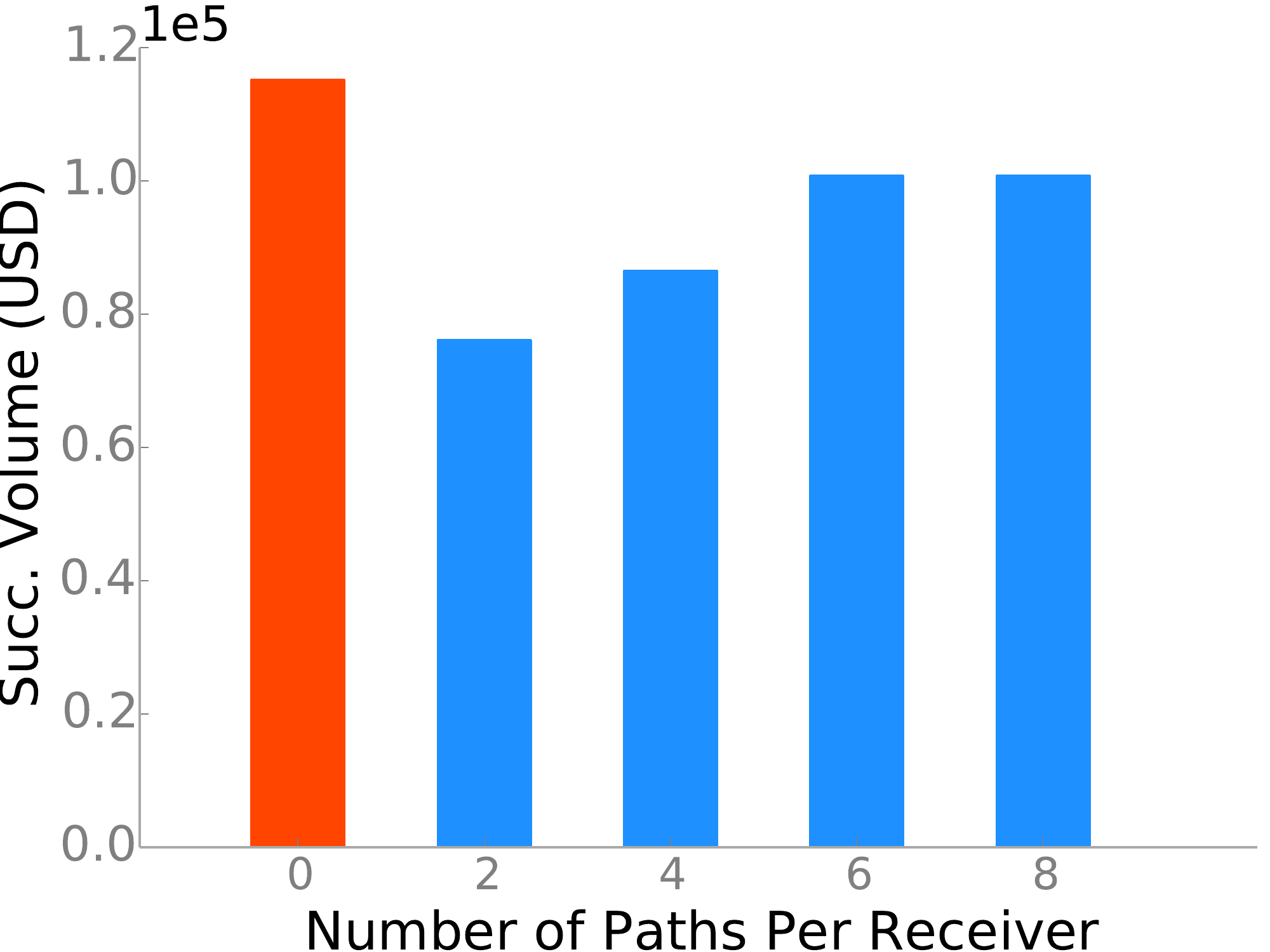}
         \caption{Success volume of mice payments}
     \end{subfigure}
     ~
     \begin{subfigure}[t]{0.485\linewidth}
         \centering
         \includegraphics[width=\textwidth]{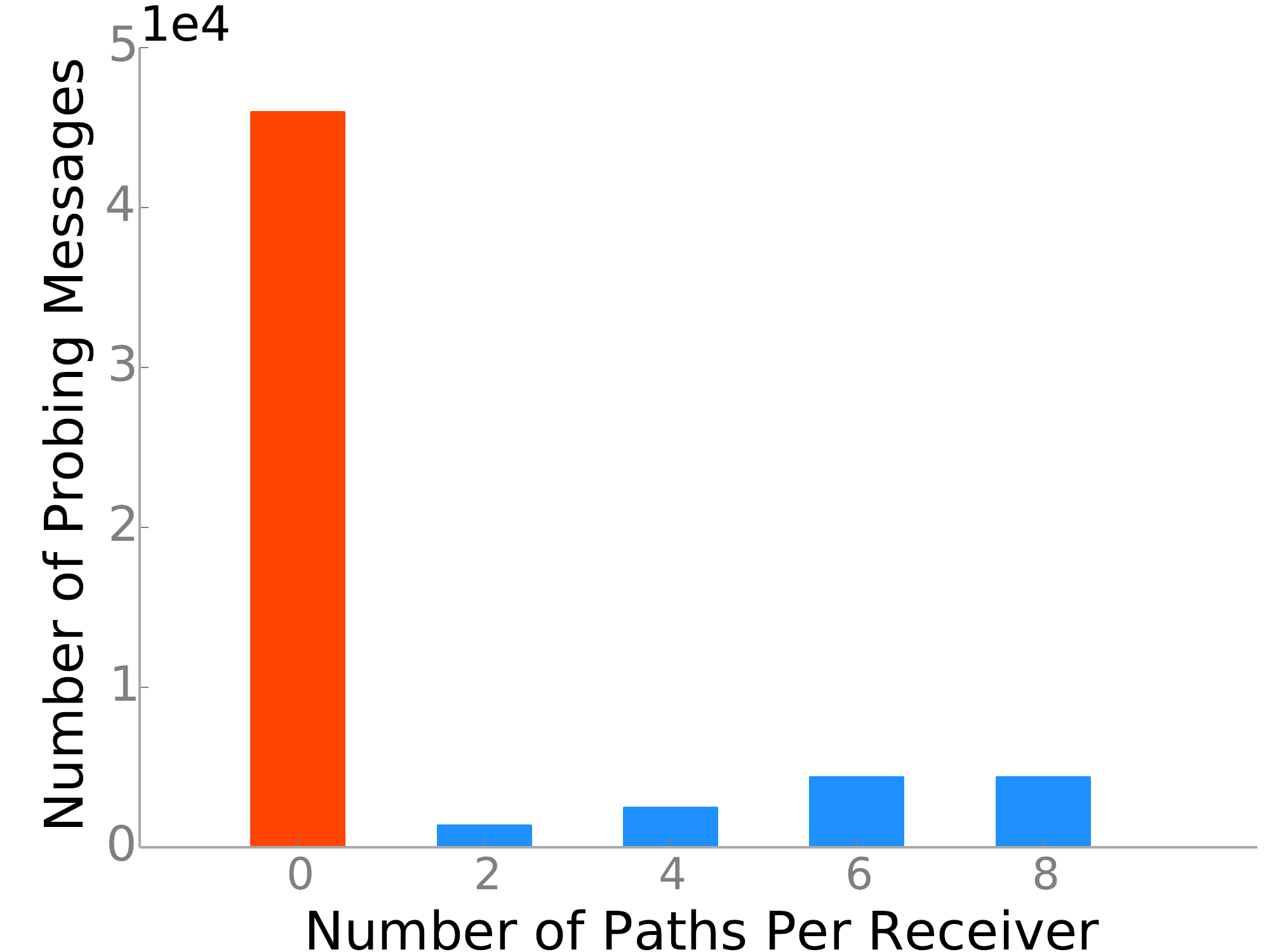}
         \caption{Probing overhead of mice payments}
     \end{subfigure}%
     \vspace{-0.2cm}
     \caption{Impact of number of paths per receiver $m$ for mice payment
     routing in \sys. Here \sys routes mice payments in the same way as
     elephant payments when $m=0$.}
     \label{fig:ripple-cache}
\end{figure}

\parabf{Impact of number of paths per receiver for mice routing.} 
We now investigate the benefit of using just a few shortest paths per receiver 
in routing mice payments. 
We only show results with the Ripple trace for brevity since results with
Lightning trace show similar trends. 
Here we vary $m$, the number of paths for a receiver in the routing table for
mice payments.
The case with $m=0$ represents the performance upperbound when we route mice
payments in the same way as elephant payments in \cref{sec:elephants}, which
clearly offers the best performance in success volume.
Figure~\ref{fig:ripple-cache}(a) shows that just a few paths per receiver 
leads to fairly good performance compared to routing them as elephants: the gap
is within 15\% with $m=6$.
The performance of \sys stabilizes when $m$ exceeds 6. 
Figure~\ref{fig:ripple-cache}(b) shows that using a few routes achieves
at least $\sim$12x less probing overhead. 
These results confirm that our mice payment routing design is effective in
reducing probing overhead while ensuring satisfactory performance.

\section{Testbed Evaluation}
\label{sec:testbed}

We conduct testbed evaluation to further investigate \sys's design.

\begin{figure*}[t]
\vspace{-0.3cm}
     \centering
\begin{subfigure}[t]{0.24\linewidth}
         \centering
         \includegraphics[width=\textwidth]{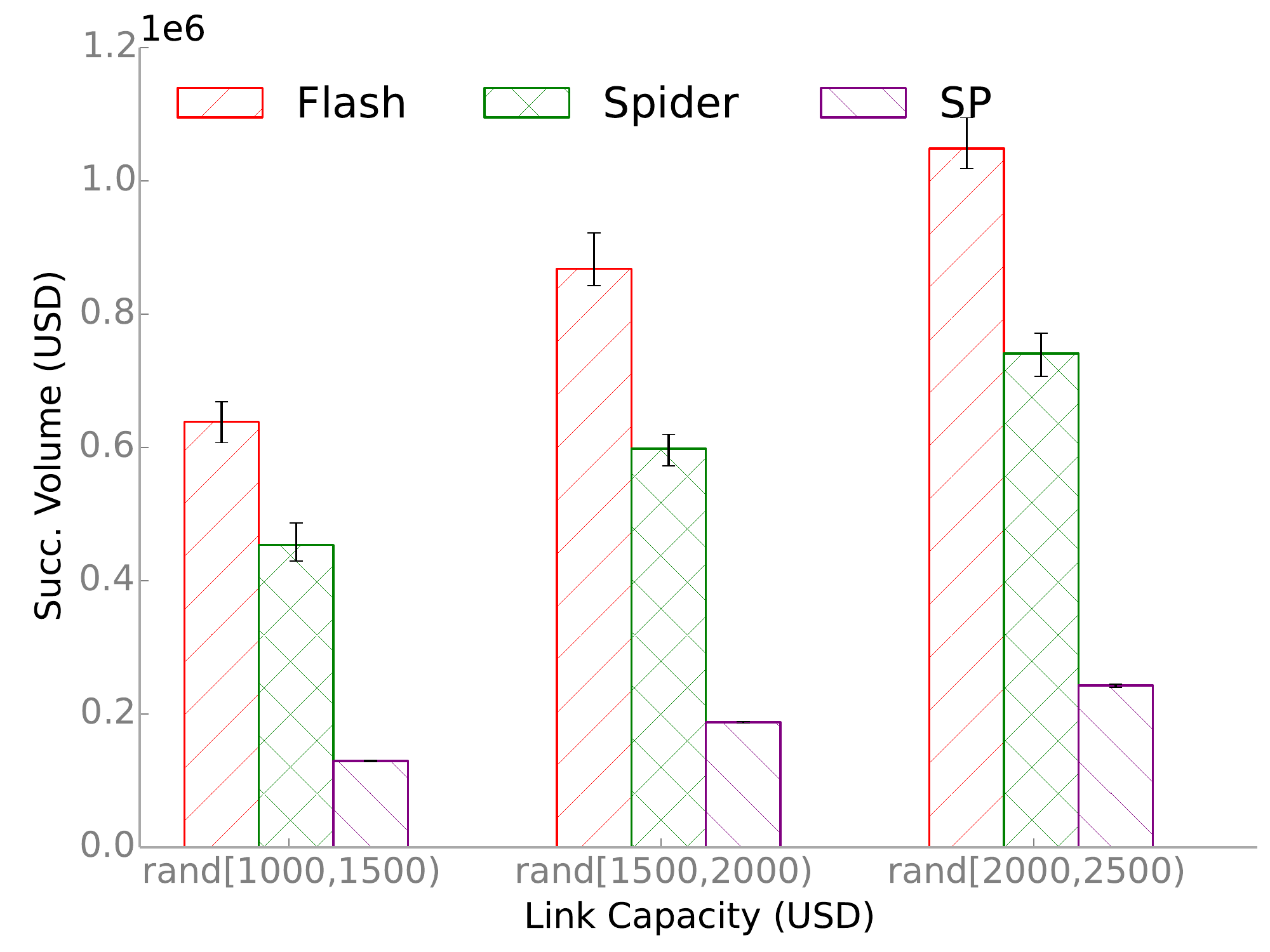}
         \caption{Success volume}
         \label{fig:tb-vol-50}
     \end{subfigure}
     \begin{subfigure}[t]{0.24\linewidth}
         \centering
         \includegraphics[width=\textwidth]{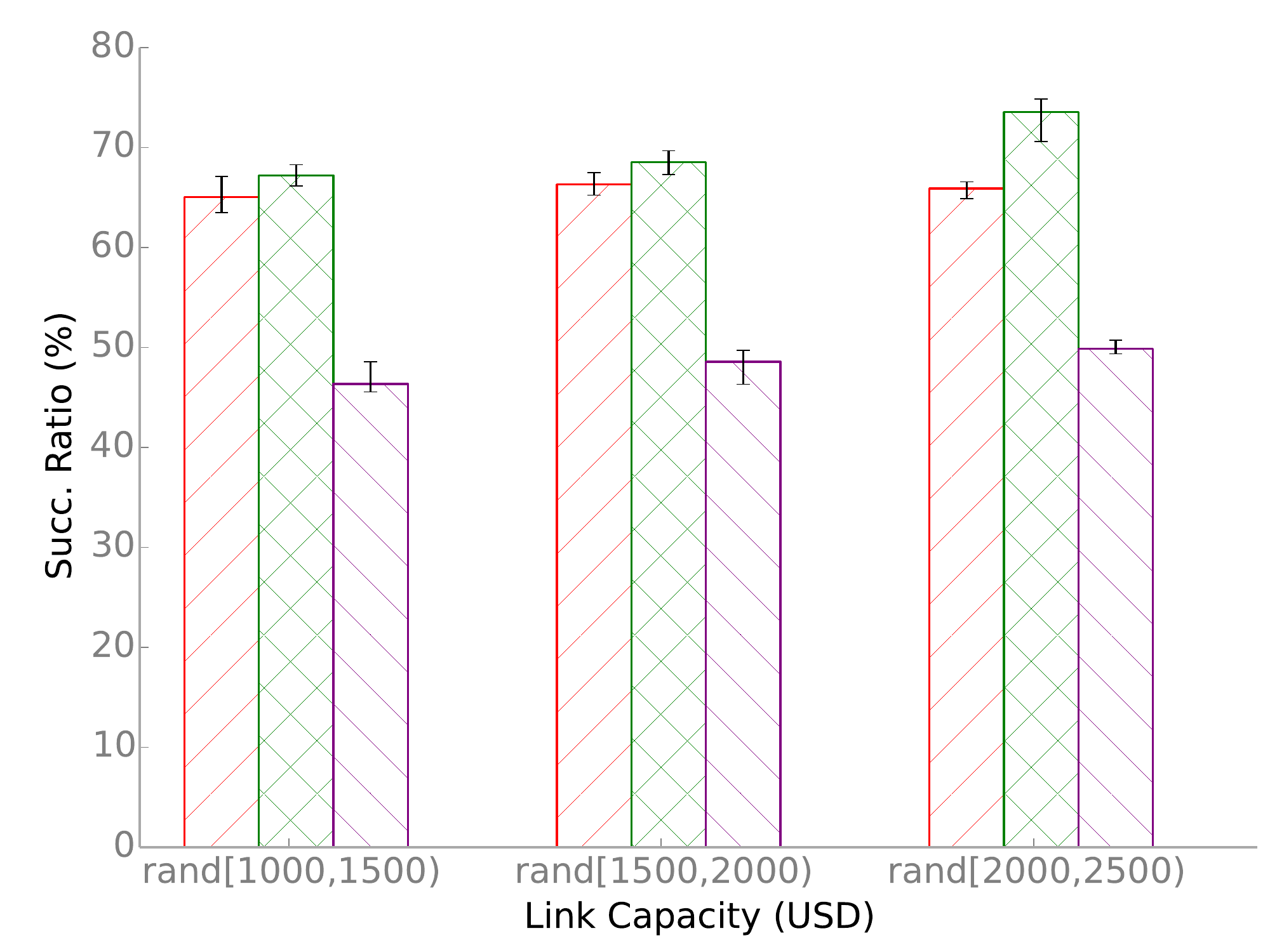}
         \caption{Success ratio}
         \label{fig:tb-ratio-50}
     \end{subfigure}        
     \begin{subfigure}[t]{0.24\linewidth}
         \centering
         \includegraphics[width=\textwidth]{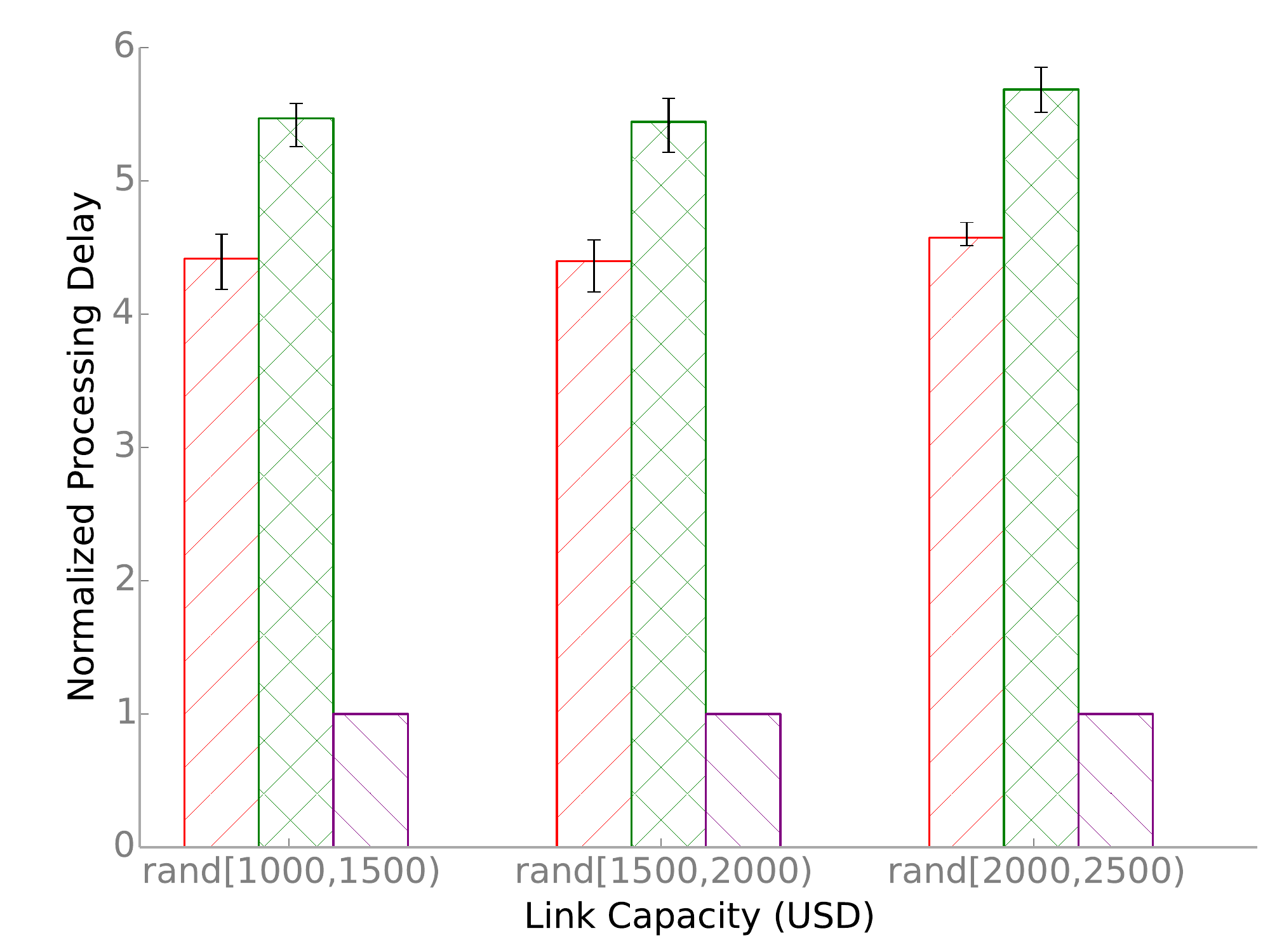}
         \caption{Overall processing delay}
         \label{fig:tb-ovh-sum-50}
     \end{subfigure}%
     \begin{subfigure}[t]{0.24\linewidth}
         \centering
         \includegraphics[width=\textwidth]{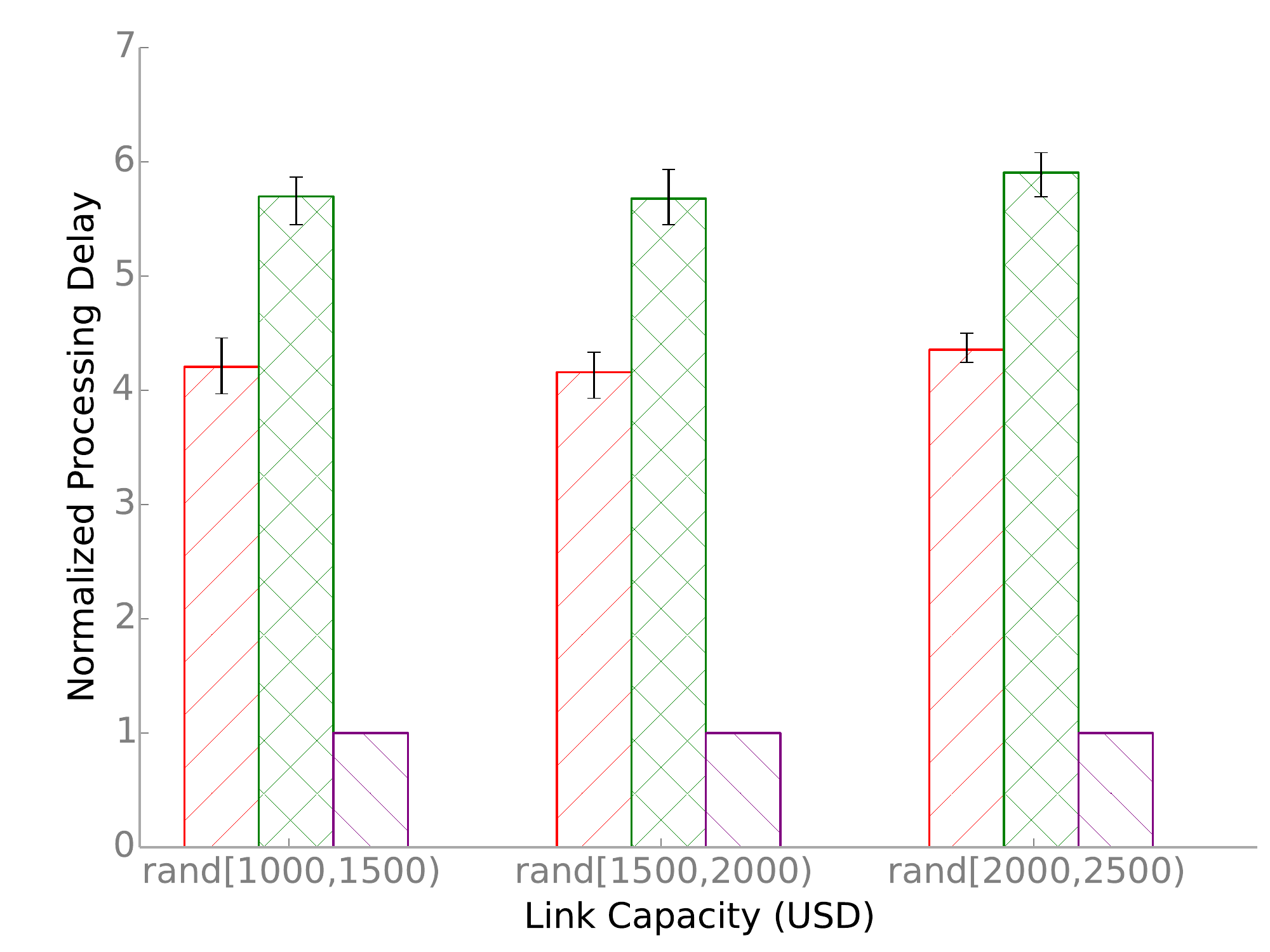}
         \caption{Mice processing delay}
         \label{fig:tb-ovh-short-50}
     \end{subfigure}%
     \vspace{-0.2cm}
     \caption{Testbed experiment results of the 50-node network.}
     \label{fig:ripple-50}
 \end{figure*}

\begin{figure*}[t]
\vspace{-0.3cm}
     \centering
      \begin{subfigure}[t]{0.24\linewidth}
         \centering
         \includegraphics[width=\textwidth]{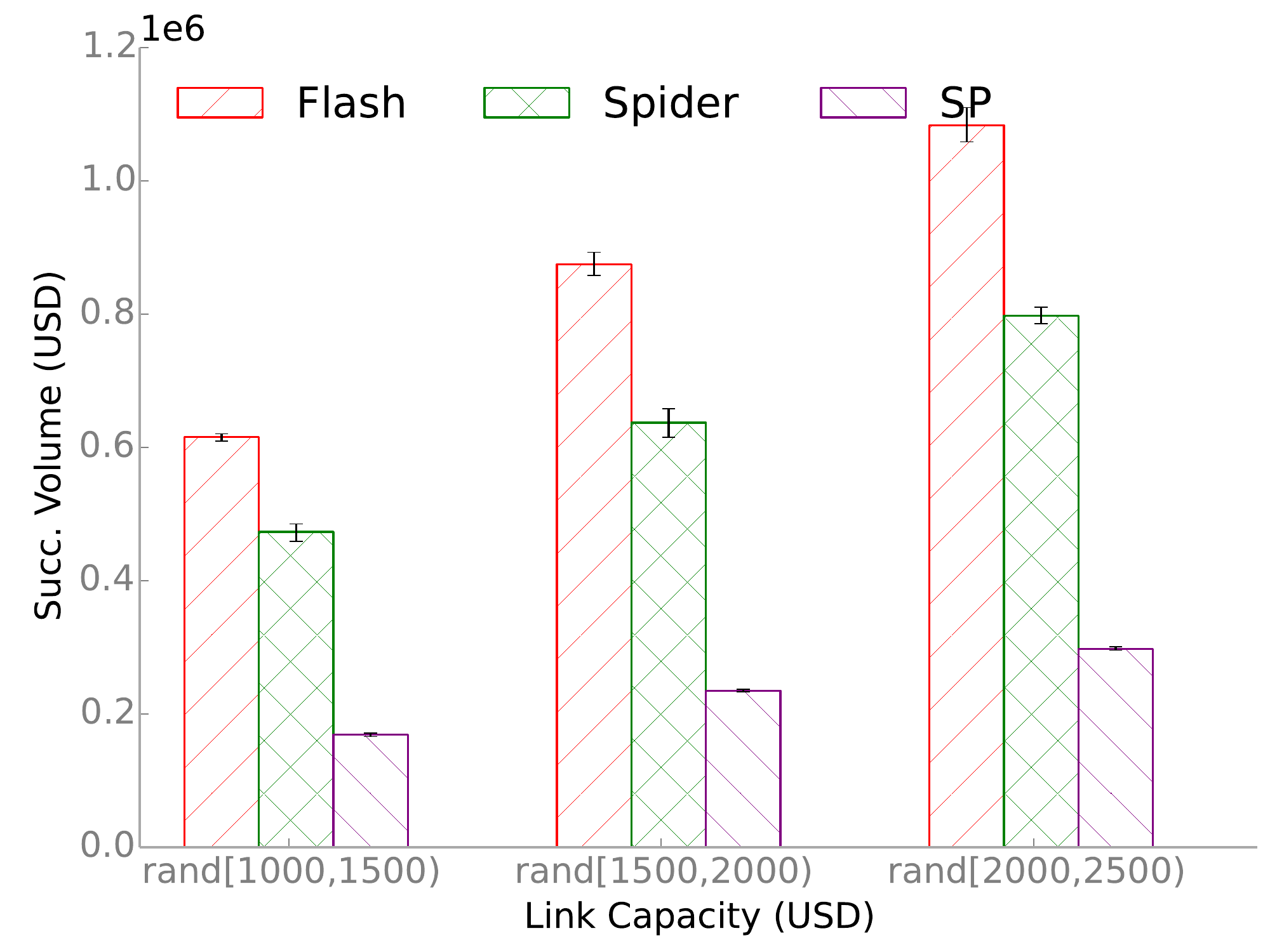}
         \caption{Success volume}
         \label{fig:tb-vol-100}
     \end{subfigure}
     \begin{subfigure}[t]{0.24\linewidth}
         \centering
         \includegraphics[width=\textwidth]{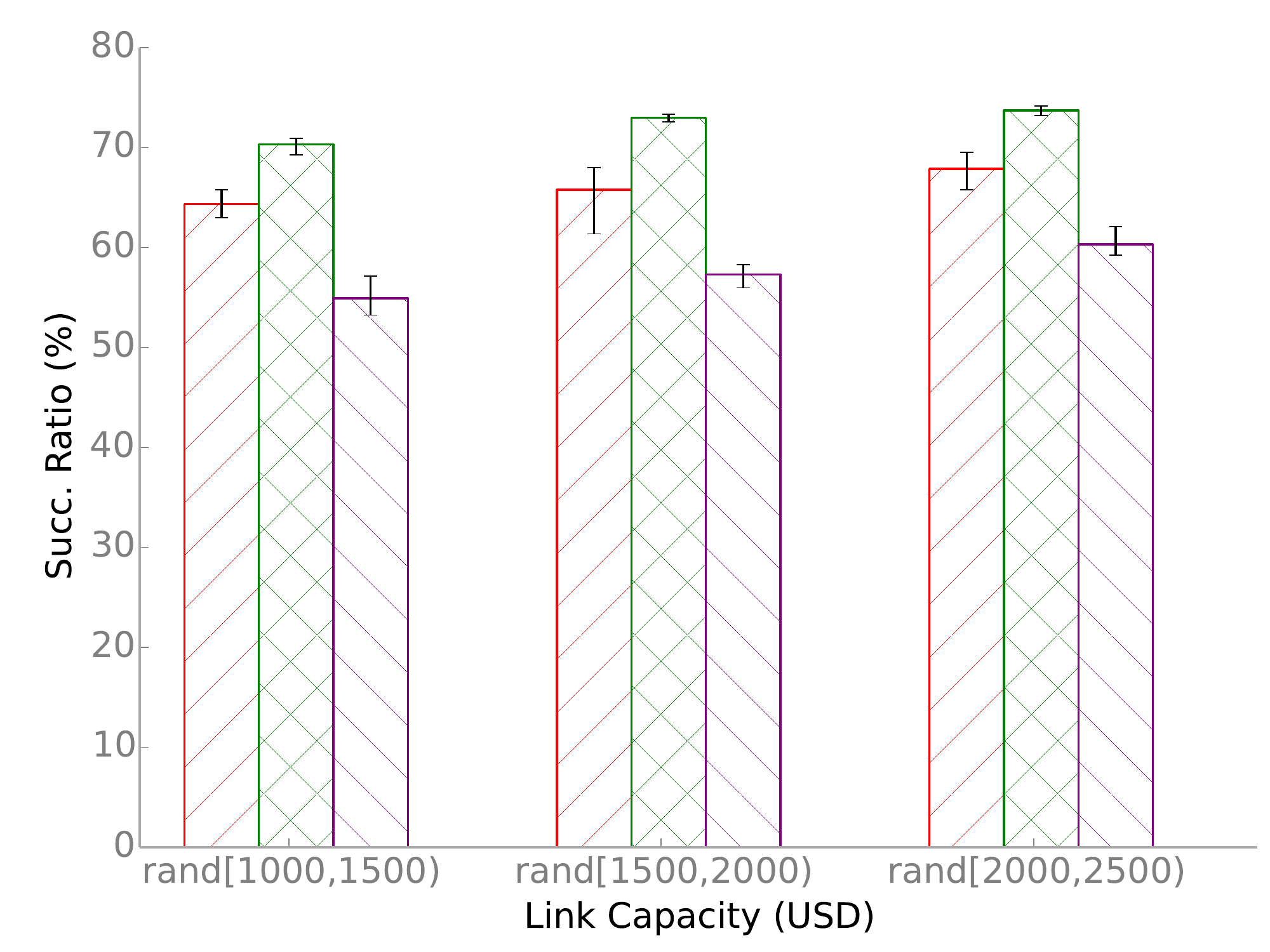}
         \caption{Success ratio}
         \label{fig:tb-ratio-100}
     \end{subfigure}
     \begin{subfigure}[t]{0.24\linewidth}
         \centering
         \includegraphics[width=\textwidth]{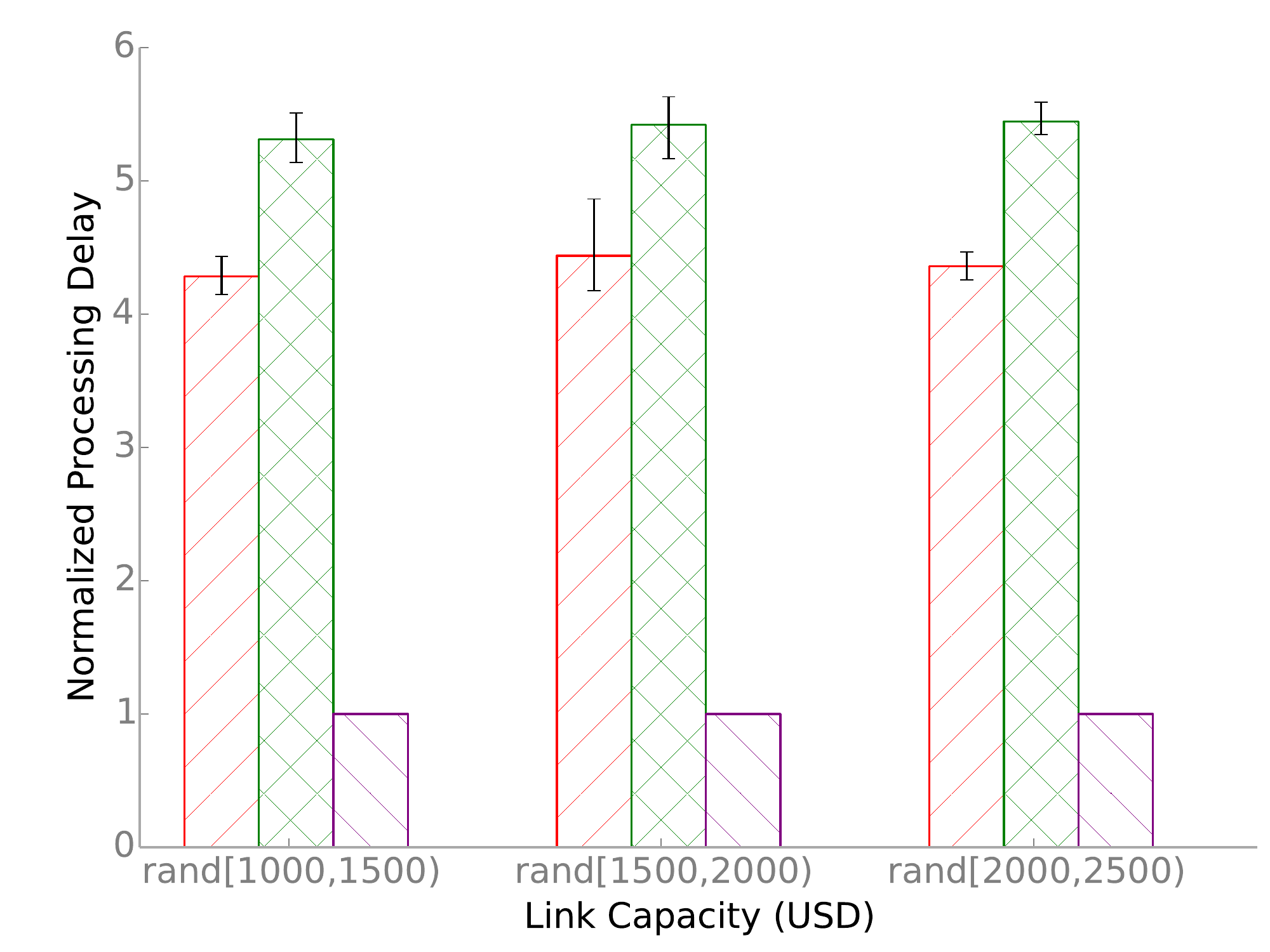}
         \caption{Overall processing delay}
         \label{fig:tb-ovh-sum-100}
     \end{subfigure}%
     \begin{subfigure}[t]{0.24\linewidth}
         \centering
         \includegraphics[width=\textwidth]{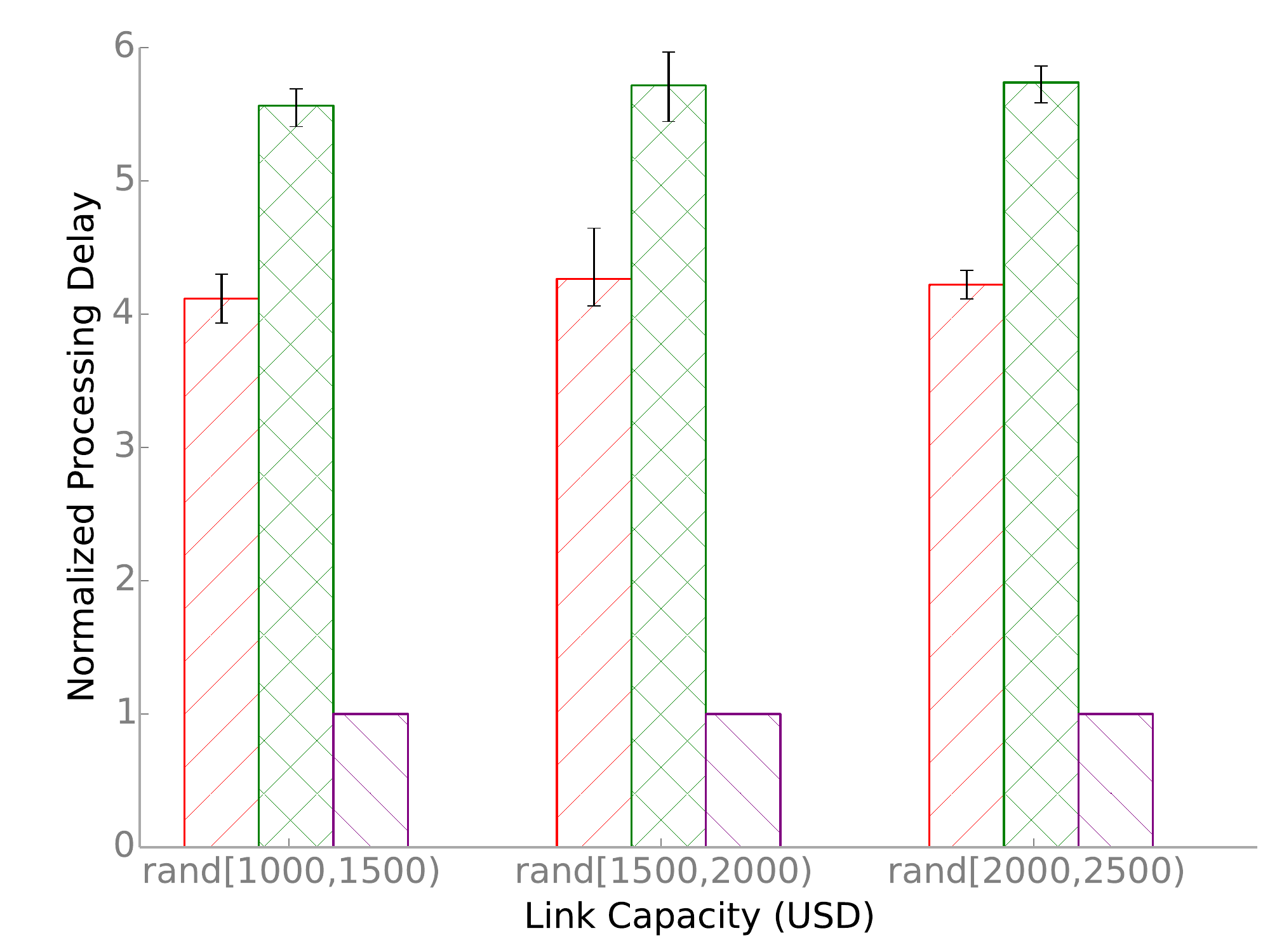}
         \caption{Mice processing delay}
         \label{fig:tb-ovh-short-100}
     \end{subfigure}%
     \vspace{-0.2cm}
     \caption{Testbed experiment results of the 100-node network.}
     \label{fig:ripple-100}
 \end{figure*}

\subsection{Implementation}
\label{sec:implementation}

We start by describing our prototype implementation. 

\parabf{Overview.} % mundane stuff, topo, connectivity, etc.
Since we focus on routing, we take a minimalist approach and build a
simplified prototype offchain routing system without mechanisms
such as the gossiping protocols for topology maintenance and HTLC for
security. We use Golang to implement the prototype with TCP for network
communication. 
The prototype reads the network topology from a local file at launch time. Upon
seeing a new transaction, it runs the routing algorithm and send it out. 

Most importantly, we implement an offchain routing protocol in our prototype
to realize three essential functions required by any routing algorithm:
source routing, probing, and atomic
payment processing. We describe the details of them in the following.

\parabf{Source routing}
is the basic service of an offchain network since the probing process and
payment routing happen over a specified path of multiple hops in the overlay
network. 
We implement a simple source routing scheme by embedding the complete path into
every message a sender initiates. 
Table~\ref{tab:impl_tab} shows the message format used in our prototype, where
the {\tt Path} field contains the path information. Upon receiving a message a
node parses this field and relays to the next-hop after necessary processing
as indicated in the {\tt Type} field.

\begin{table}[t]
\small
\begin{tabular}{c|c}
\hline
\centering
Field & Description \\ \hline
{\tt TransID} & A unique ID of a (partial) payment \\
{\tt Type} & Message type \\
{\tt Path} & Path of this message \\
{\tt Capacity} & Probed channel capacity \\
{\tt Commit} & Committed amount of funds for this payment \\ \hline
\end{tabular}
\caption{Message format for source routing in our prototype.}
\label{tab:impl_tab}
\end{table}

\parabf{Probing}
is needed for offchain routing algorithms to collect the ever-changing channel
balance. A node initiates probing by constructing a {\tt PROBE} message with
the path it is interested in. The intermediate nodes append the {\tt Capacity}
field in the message with their current balance. To return the probed
information, the receiver modifies the message type to {\tt PROBE\_ACK},
replaces the {\tt Path} field with the reversed version of the forward path,
and sends it back all the way to the sender.

\parabf{Atomic payment processing.}
Last but not least, we implement a two-phase commit protocol to realize atomic
payment processing without complex security mechanisms like HTLC 
\cite{poon2016bitcoin}. 
This is necessary for two reasons. 
First, due to network dynamics it is possible that a payment fails
on its path because the balance of some channel has changed after it was
last probed by the sender. 
Thus confirmation is required for the sender to know the status of the payment
and ensure the atomicity of balance update on the path. 
Second and more interestingly, with multipath routing a payment is successful if
and only if all its sub-payments are successful \cite{lightningnetwork,raiden}. 
This necessitates the need of two-phase commit from distributed systems, where
the protocol only commits
the payment when all its sub-payments have been confirmed on their paths. 

Specifically, our two-phase commit protocol works as follows for the general
case of multipath routing. In the first phase, the sender prepares a {\tt
COMMIT} message for each of the sub-payment and sends them out. An
intermediate node determines if its current balance can handle this
sub-payment. If yes, it decreases its balance by the volume specified in the
{\tt COMMIT} message and forwards it to the next hop. The receiver constructs
a {\tt COMMIT\_ACK} message by adding the success information in the payload
and reversing the path. The sender recognizes this sub-payment to be
successful upon receiving the {\tt COMMIT\_ACK}. In case an intermediate node
does not have enough balance, it constructs a {\tt COMMIT\_NACK} with the
reversed path and immediately sends it back to its previous hop. 
The sender recognizes the sub-payment to be
failed afterwards.

After results of all sub-payments are back, the protocol enters the second
phase. 
When all sub-payments are successful, the sender sends a {\tt CONFIRM} message
for each sub-payment along their paths. 
The intermediate nodes simply relay the {\tt CONFIRM} message. 
The receiver would send a {\tt CONFIRM\_ACK} along the reverse path back to the
sender. 
Now each intermediate node processes {\tt CONFIRM\_ACK} by adding the committed
funds of this sub-payment to the channel in the reverse direction, in order to
make the bidirectional channel balances consistent.
With all {\tt CONFIRM\_ACK} received the sender considers this payment
successful. 
In case at least one sub-payment is unsuccessful in the first commit phase, the
sender sends a {\tt REVERSE} message for each sub-payment. 
All intermediate nodes then add back the committed funds to the channel in the
forward path, and the receiver sends a corresponding {\tt REVERSE\_ACK} to
indicate that everyone has been informed.

\subsection{Experiment Setup}
\label{sec:testbed_setup}

We now present the setup of our testbed and experimental results.

Our evaluation is conducted on a server machine with a 10-core
Intel E5-2640v4 CPU and 64GB DDR4-2400 memory.  
For simplicity we represent each node of an offchain network as a single process
running our Golang prototype.
Each process is bound to a unique ip address and port number tuple.

We implement our routing algorithms described in \cref{sec:design} in our
prototype. We also implement two baseline routing algorithms: Spider as in 
\cite{sivaraman2018routing} and a simple shortest path scheme (denoted as SP) as
described in \cref{sec:sim_setup}.

The network topology follows the Watts Strogatz graph \cite{watts1998collective}. 
We generate two topologies each with 50 and 100 nodes, respectively. 
The capacity of each channel is set randomly from an interval which varies from
$[\$1000,\$1500)$, $[\$1500,\$2000)$, to $[\$2000,\$2500)$. 
We generate 10,000 transactions whose volume follows the Ripple trace and
randomly select the sender-receiver pairs\footnote{We ensure there exists at least one path from sender to receiver.}.
For \sys the payment size threshold is set such that 90\% of transactions are
mice, the number of paths for elephant routing $k$ is 20, and the number of
shortest paths for mice routing $m$ is 4.
Spider uses 4 edge-disjoint shortest paths as proposed in 
\cite{sivaraman2018routing}. 
Each scheme is evaluated in 5 independent runs. Results are shown with
min-mean-max bars.
\subsection{Results}
\label{sec:testbed_results}

We first look at performance. 
We can observe from Figures~\ref{fig:tb-vol-50} and
\ref{fig:tb-vol-100} that the success volume of \sys is much larger than Spider, 42.5\%
and 34.4\% on average for the 50-node and 100-node topologies, respectively. This demonstrates the
effectiveness of our routing algorithms which select a good set of paths to
improve throughput. 
As shown in Figures~\ref{fig:tb-ratio-50} and
\ref{fig:tb-ratio-100},  \sys's success ratio is slightly worse (5.6\% and 8.8\%
on average, respectively)
than Spider and is better (36.3\% and 14.8\% on average, respectively) than
SP. 
The reason \sys has lower success ratio than Spider is that Spider uses
waterfilling to balance the utilization of multiple paths and creates better
chances for mice payments to go through. 
\sys does not consider load balance in its design.

Next we investigate overhead. Instead of messaging overhead in our testbed
evaluation we measure the average processing delay of a transaction in our
prototype as the metric of overhead. 
We normalize the results by the average processing delay of SP, the
simplest baseline algorithm.
From Figures~\ref{fig:tb-ovh-sum-50} and \ref{fig:tb-ovh-sum-100}, we can see that 
\sys's processing delay is on average 19.4\% and 19.2\% smaller than
Spider for the 50-node and 100-node topologies, respectively. 
Further, we look at the processing delay of mice payments which generally
require faster payment settlement time. As plotted in Figure~\ref{fig:tb-ovh-short-50} 
and \ref{fig:tb-ovh-short-100}, \sys is on average 26.4\% and 26\% faster than
Spider in the two topologies, respectively. 
This verifies our mice payment routing algorithm reduces the probing overhead
and thus the processing delay significantly.

\section{Related Work}
\label{sec:relatedwork}

Offchain routing emerges only recently in 2016.
The first offchain routing algorithm is proposed in the design draft of
Lightning
network \cite{poon2016bitcoin}. It routes payments to paths using a BGP-like
system and maintains a global routing table. To minimize the routing state,
Flare \cite{prihodko2016flare} proposes that nodes only maintain neighbors
within a certain hop distance. When routing a payment, the sender exchanges
the neighbor information with the receiver to construct complete paths. Besides,
each node finds some random beacon nodes to supplement its view of the network.

To further reduce the message overhead in path finding, SilentWhispers
\cite{moreno2017silentwhispers} utilizes landmark-centered routing. It performs
periodic Breadth-First-Search to find the shortest path from the landmarks to the
sender and receiver. All paths need to go through the landmarks, which
makes some paths unnecessarily long. SpeedyMurmurs \cite{roos2018settling}
proposes embedding-based routing to assign coordinates to nodes and find
shortcuts that reduce the average path lengths.

The above routing algorithms fall into static routing, which does
not consider payment channel dynamics and leads to poor throughput performance.
Revive \cite{khalil2017revive} and Spider \cite{sivaraman2018routing} take the
dynamic channel balances into consideration and propose centralized offline
routing algorithms to maximize the throughput or success volume of
payments.
As we discussed centralized schemes have high probing overhead and do not work
for decentralized offchain networks.

Compared to exiting work, \sys is the first solution that considers the
characteristics of payments in offchain network in order to achieve a
better balance between the path optimality and probing overhead.
\sys's approach of differentiating elephant and mice payments are akin to past
work on flow scheduling in datacenter networks (DCNs), such as Hedera
\cite{ARRH10} and DevoFlow \cite{CMTY11}. Other effective approaches for DCNs,
such as congestion aware load balancing \cite{AEDV14,WXNH16} and fine-grained
routing \cite{VPAT17,HRAF15,PBS17}, may also provide insights for offchain
routing solutions. The key differences are that, an offchain network topology
is highly irregular while a DCN topology is usually a Clos, and the channel
balance is highly fluctuating while the link capacity is fixed and abundant in
a DCN. We believe how to learn from these proven ideas in DCN for better
offchain routing designs would be an interesting direction of future work with
much potential.

\section{Conclusion}
\label{sec:conclusion}

We presented \sys, a new routing solution that efficiently delivers payments
over offchain networks.
By studying the characteristics of payments in real offchain
networks, we find that payment sizes are heavy-tailed, and most payment are
recurring.
\sys thus differentiates the treatment of elephant and mice payments. It uses
a modified max-flow algorithm to provide elephant payments with sufficient
path capacity, and routes mice payments by a routing table with just a few
shortest paths to reduce probing overhead. Through trace-driven simulations
and prototype implementation, we demonstrated that \sys significantly
outperforms existing solutions especially on success volume, while maintaining
low probing overhead.

% \clearpage
\bibliographystyle{abbrv}
\bibliography{reference}

\begin{thebibliography}{10}

\bibitem{amp}
{Atomic Multi-path Payment}.
\newblock
  \url{https://lists.linuxfoundation.org/pipermail/lightning-dev/2018-February/000993.html}.

\bibitem{clightning}
{c-lightning Daemon}.
\newblock
  \url{https://github.com/ElementsProject/lightning/tree/master/lightningd}.

\bibitem{lightingnetworkcode}
{Lightning Network Daemon}.
\newblock \url{https://github.com/lightningnetwork/lnd}.

\bibitem{networkx}
{NetworkX}.
\newblock \url{https://networkx.github.io/}.

\bibitem{raidennetworkcode}
{Raiden Network Daemon}.
\newblock \url{https://github.com/raiden-network/raiden}.

\bibitem{ripple-trace}
Ripple transaction trace.
\newblock \url{https://crysp.uwaterloo.ca/software/speedymurmurs/}, 2018.

\bibitem{bitcoin}
Bitcoin.
\newblock \url{https://bitcoin.org/en/}, 2019.

\bibitem{lightning-stats}
{Real-Time Lightning Network Statistics}.
\newblock \url{https://1ml.com/statistics}, January 2019.

\bibitem{ripple}
Ripple.
\newblock \url{https://ripple.com/}, 2019.

\bibitem{lightningnetwork}
{The Lightning Network}.
\newblock \url{https://lightning.network/}, 2019.

\bibitem{raiden}
{The Raiden Network}.
\newblock \url{https://raiden.network/}, 2019.

\bibitem{tps}
{Transaction Rate of Bitcoin}.
\newblock \url{https://www.blockchain.com/en/charts/transactions-per-second},
  2019.

\bibitem{ARRH10}
M.~Al-Fares, S.~Radhakrishnan, B.~Raghavan, N.~Huang, and A.~Vahdat.
\newblock {Hedera: Dynamic Flow Scheduling for Data Center Networks}.
\newblock In {\em Proc.~USENIX NSDI}, 2010.

\bibitem{AEDV14}
M.~Alizadeh, T.~Edsall, S.~Dharmapurikar, R.~Vaidyanathan, K.~Chu,
  A.~Fingerhut, V.~T. Lam, F.~Matus, R.~Pan, N.~Yadav, and G.~Varghese.
\newblock {CONGA: Distributed Congestion-Aware Load Balancing for Datacenters}.
\newblock In {\em Proc.~ACM SIGCOMM}, 2014.

\bibitem{CLRS09}
T.~H. Cormen, C.~E. Leiserson, R.~L. Rivest, and C.~Stein.
\newblock {\em {Introduction to Algorithms}}.
\newblock MIT Press, 2009.

\bibitem{CMTY11}
A.~R. Curtis, J.~C. Mogul, J.~Tourrilhes, P.~Yalagandula, P.~Sharma, and
  S.~Banerjee.
\newblock {DevoFlow: Scaling Flow Management for High-performance Networks}.
\newblock In {\em Proc.~ACM Sigcomm}, 2011.

\bibitem{ford1956maximal}
L.~R. Ford and D.~R. Fulkerson.
\newblock Maximal flow through a network.
\newblock {\em Canadian journal of Mathematics}, 8:399--404, 1956.

\bibitem{gilad2017algorand}
Y.~Gilad, R.~Hemo, S.~Micali, G.~Vlachos, and N.~Zeldovich.
\newblock Algorand: Scaling byzantine agreements for cryptocurrencies.
\newblock In {\em Proc.~ACM SOSP}, 2017.

\bibitem{HRAF15}
K.~He, E.~Rozner, K.~Agarwal, W.~Felter, J.~Carter, and A.~Akella.
\newblock {Presto: Edge-based Load Balancing for Fast Datacenter Networks}.
\newblock In {\em Proc.~ACM SIGCOMM}, 2015.

\bibitem{HKMZ13}
C.-Y. Hong, S.~Kandula, R.~Mahajan, M.~Zhang, V.~Gill, M.~Nanduri, and
  R.~Wattenhofer.
\newblock Achieving high utilization with software-driven {WAN}.
\newblock In {\em Proc.~ACM SIGCOMM}, 2013.

\bibitem{JKMO13}
S.~Jain, A.~Kumar, S.~Mandal, J.~Ong, L.~Poutievski, A.~Singh, S.~Venkata,
  J.~Wanderer, J.~Zhou, M.~Zhu, J.~Zolla, U.~H\"{o}lzle, S.~Stuart, and
  A.~Vahdat.
\newblock {B4}: Experience with a globally-deployed software defined {WAN}.
\newblock In {\em Proc.~ACM SIGCOMM}, 2013.

\bibitem{khalil2017revive}
R.~Khalil and A.~Gervais.
\newblock Revive: Rebalancing off-blockchain payment networks.
\newblock In {\em Proc.~ACM CCS}, 2017.

\bibitem{luu2016secure}
L.~Luu, V.~Narayanan, C.~Zheng, K.~Baweja, S.~Gilbert, and P.~Saxena.
\newblock A secure sharding protocol for open blockchains.
\newblock In {\em Proc.~ACM CCS}, 2016.

\bibitem{moreno2017silentwhispers}
P.~Moreno-Sanchez, A.~Kate, and M.~Maffei.
\newblock Silentwhispers: Enforcing security and privacy in decentralized
  credit networks.
\newblock In {\em 24th Network and Distributed System Security Symposium (NDSS
  2018)}, 2017.

\bibitem{nakamoto2008bitcoin}
S.~Nakamoto.
\newblock Bitcoin: A peer-to-peer electronic cash system.
\newblock {\em Technical Report, https://bitcoin.org/bitcoin.pdf}, 2008.

\bibitem{PBS17}
J.~Perry, H.~Balakrishnan, and D.~Shah.
\newblock {Flowtune: Flowlet Control for Datacenter Networks}.
\newblock In {\em Proc.~USENIX NSDI}, 2017.

\bibitem{poon2016bitcoin}
J.~Poon and T.~Dryja.
\newblock The bitcoin lightning network: Scalable off-chain instant payments.
\newblock {\em Technical Report,
  https://lightning.network/lightning-network-paper.pdf}, 2016.

\bibitem{prihodko2016flare}
P.~Prihodko, S.~Zhigulin, M.~Sahno, A.~Ostrovskiy, and O.~Osuntokun.
\newblock Flare: An approach to routing in lightning network.
\newblock {\em White Paper
  (bitfury.com/content/5-white-papers-research/whitepaper\_flare\_an\_approach\_to\_routing\_in\_lightning\_n
  etwork\_7\_7\_2016. pdf)}, 2016.

\bibitem{roos2018settling}
S.~Roos, P.~Moreno-Sanchez, A.~Kate, and I.~Goldberg.
\newblock {Settling Payments Fast and Private: Efficient Decentralized Routing
  for Path-Based Transactions}.
\newblock In {\em Proc.~NDSS}, 2018.

\bibitem{sivaraman2018routing}
V.~Sivaraman, S.~B. Venkatakrishnan, M.~Alizadeh, G.~Fanti, and P.~Viswanath.
\newblock Routing cryptocurrency with the spider network.
\newblock 2018.

\bibitem{VPAT17}
E.~Vanini, R.~Pan, M.~Alizadeh, P.~Taheri, and T.~Edsall.
\newblock {Let It Flow: Resilient Asymmetric Load Balancing with Flowlet
  Switching}.
\newblock In {\em Proc.~USENIX NSDI}, 2017.

\bibitem{WW19}
J.~Wang and H.~Wang.
\newblock {Monoxide: Scale Out Blockchain with Asynchronized Consensus Zones}.
\newblock In {\em Proc.~USENIX NSDI}, 2019.

\bibitem{WXNH16}
P.~Wang, H.~Xu, Z.~Niu, D.~Han, and Y.~Xiong.
\newblock {Expeditus: Congestion-aware Load Balancing in Clos Data Center
  Networks}.
\newblock In {\em Proc.~ACM SoCC}, 2016.

\bibitem{watts1998collective}
D.~J. Watts and S.~H. Strogatz.
\newblock Collective dynamics of `small-world'networks.
\newblock {\em nature}, 393(6684):440, 1998.

\bibitem{wood2017ethereum}
G.~Wood.
\newblock Ethereum: a secure decentralized transaction ledger, 2014.

\bibitem{yen1971finding}
J.~Y. Yen.
\newblock Finding the k shortest loopless paths in a network.
\newblock {\em management Science}, 17(11):712--716, 1971.

\bibitem{zamani2018rapidchain}
M.~Zamani, M.~Movahedi, and M.~Raykova.
\newblock Rapidchain: scaling blockchain via full sharding.
\newblock In {\em Proceedings of the 2018 ACM SIGSAC Conference on Computer and
  Communications Security}, pages 931--948. ACM, 2018.

\end{thebibliography}

\end{document}